\documentclass[11pt,a4paper,english]{article}
\usepackage[T1]{fontenc}
\usepackage[utf8]{inputenc}
\usepackage{babel}

\usepackage{amsmath}
\usepackage{amsthm}
\usepackage{amsfonts}
\usepackage{amssymb}

\usepackage{graphicx}
\usepackage{mathtools}
\usepackage{comment}
\usepackage{lipsum}
\usepackage{caption} 
\usepackage{xcolor}
\usepackage{dirtytalk}
\usepackage{natbib}
\usepackage{hyperref}

\title{The PEAL Method: a mathematical framework to streamline securitization structuring}
\author{%
  Andrea Pinto\\
  \small Private Practice\\
  \small Neuture.ai \\
  \small\texttt{andrea.pinto@neuture.ai}
  \and
  Antonio Scala\\
  \small Institute for Complex Systems\\
  \small CNR\\
  \small\texttt{antonio.scala@cnr.it}
}

\begin{document}
  \maketitle

  \begin{abstract}
Securitization is a financial process where the cash flows of income-generating assets are sold to institutional investors as securities, liquidating illiquid assets. This practice presents persistent challenges due to the absence of a comprehensive mathematical framework for structuring asset-backed securities. While existing literature provides technical analysis of credit risk modeling, there remains a need for a definitive framework detailing the allocation of the inbound cash flows to the outbound positions. To fill this gap, we introduce the PEAL Method: a 10-step mathematical framework to streamline the securitization structuring across all time periods.

The PEAL Method offers a rigorous and versatile approach, allowing practitioners to structure various types of securitizations, including those with complex vertical positions. By employing standardized equations, it facilitates the delineation of payment priorities and enhances risk characterization for both the asset and the liability sides throughout the securitization life cycle.

In addition to its technical contributions, the PEAL Method aims to elevate industry standards by addressing longstanding challenges in securitization. By providing detailed information to investors and enabling transparent risk profile comparisons, it promotes market transparency and enables stronger regulatory oversight.

In summary, the PEAL Method represents a significant advancement in securitization literature, offering a standardized framework for precision and efficiency in structuring transactions. Its adoption has the potential to drive innovation and enhance risk management practices in the securitization market.
  \end{abstract}
  \newpage

  \tableofcontents\newpage
  
\section{Unboxing securitizations} \label{Sec: C1 - unboxing}

\subsection{Introduction} \label{Subsec: C1 - intro}
Securitization is a financial process where the cash flows of income-generating assets are sold to institutional investors as securities, liquidating illiquid assets. Despite securitization modelling complexity, the literature addressing the requisite mathematical modeling is sparse. While there are a few notable exceptions - such as the two excellent books describing in detail the issues encountered in structuring asset-backed securities \citep{bluhm2006structured,bluhm2010introduction} - a comprehensive mathematical framework describing how the inbound cash flows are allocated to the outbound positions is lacking. This paper introduces the PEAL Method, a 10-step mathematical framework designed to streamline securitization structuring while encompassing a comprehensive analysis across all time periods. It offers a standardized mathematical approach adaptable to various scenarios, ensuring compliance with regulations and enabling transparent risk assessment for both the asset and the liability sides. Inspired by Queuing Theory, which elucidates queuing system dynamics, we present securitization as a mathematical study akin to optimizing queuing processes. By establishing a link between inbound and outbound cash flows in 10 steps, our method aims to enhance efficiency and affordability in structuring securitizations, ultimately fostering transparency and regulatory compliance.

\subsection{Securitization definition}\label{Subsec: C1 - SecDef}
European Regulation No 2402/2017 (\textbf{Reg 2402}), Article 2(1), defines securitization as a \say{transaction or scheme, whereby the credit risk associated with an exposure or a pool of exposures} (\textbf{Exposures}) \say{is tranched, having all of the following characteristics: (a) payments} [to the securitization positions (\textbf{Positions}, see definition in Sec. \ref{Subsec: Step 5 - Positions Definition})] \say{in the transaction or scheme are dependent upon the performance} of the Exposures; \say{(b) the subordination of tranches determines the distribution of losses during the \textit{ongoing life} of the transaction or scheme; (c) the transaction or scheme does not create exposures which possess all of the characteristics listed in Article 147(8) of Regulation No 575/2013}.

Article 2(1)(a) requires that the payments toward the Positions be dependent upon the performance of the Exposures. Therefore a scheme or transaction is not considered a securitization if the Exposures \textit{ongoing losses} are always zero $\forall \ t$. In fact, if this were the case, there would be no credit risk to \say{tranche}, and thus the letter (b) requirement would not be satisfied. Whilst it is necessary the presence of potential losses, it is not sufficient per sé. Article 2(1)(b) additionally requires to \say{tranche the credit risk} in such a way that the losses be allocated to the different Positions so as to reflect the subordination of tranches continuously during their \emph{ongoing life}.

Thus, Reg 2402 defines securitization in mathematical terms: in order to classify a scheme or transaction as a \emph{securitization}, both the Exposures' ongoing losses and the Positions' cash flows must be calculated and jointly analyzed. To our best knowledge, there is no prior paper that has defined a clear relationship between these two elements. Indeed, we aspire to fill this gap, identifying the main inbound and outbound building-blocks needed to structure a securitization and providing a conceptual and mathematical definition of each element individually and in correlation to the others.

This paper\footnote{Unless otherwise defined all terms of this paper have the same meaning of Reg 2402.} introduces the \textbf{PEAL Method}\footnote{PEAL stands for (P)ositions, (E)vents recovery, (A)ssets, and (L)osses and represents the name of the macro building-blocks that are at the basis of the Securitization Theory.} that, correlating the inbound to the outbound cash flows, opens up the securitization black-box and enables more sophisticated and robust statistical analyses. Our mathematical approach leaves less room for possible arrangers' misjudgments, all while easing the Public Authorities supervision.

\subsection{Securitization: asset-side basic variables} \label{Subsec: C1 - Exposures Inputs}
Let the asset-side of a securitization be composed of $K \ge 1$ portfolios, and let each portfolio $k$ have a number of Exposures $N_k$. Each of the $n=1,\dots,N_k$ Exposures of portfolio $k$ is amortized over $T_{k,n}$ months, that are pooled together into the securitization at month $\hat{\tau}_k$ (where $T_{k,n}$ and $\hat{\tau}_k$ $\in N$). By exploiting these definitions, let a securitization be \emph{Basic} when all the portfolios $K$ are structured at the same initial time $\hat{\tau}_1 = \hat{\tau}_2 = \dots \hat{\tau}_K = \hat{\tau}^*$. On the other hand, let a securitization be \emph{Rolling} when at least one of the portfolios $K \geq 2$ is structured at a time $\hat{\tau}_k > \hat{\tau}^*$. We indicate by $N$ 
\begin{equation} \label{eq:N}
	N = \sum_{k=1}^{K} N_k
\end{equation}
the total number of Exposures generating inbound cash flows. If $T_k$ 
\begin{equation} \label{eq:Tk}
	T_{k} = \max_{n= 1 \dots N_k} \, T_{k,n} + \hat{\tau}_k
\end{equation}
is the duration of each portfolio $k$, then the Exposures maximum duration is
\begin{equation} \label{eq:T}
	T = \max_k \, T_k
\end{equation} 
We assume that the time when Exposures are pooled together is
\begin{equation} \label{eq:tau}
	\hat{\tau}^* = \min_k \, \hat{\tau}_k
\end{equation} 
where $\hat{\tau}^*$ is the origin of our timeline that, when not differently specified, is $\hat{\tau}^* = 0$. Then for each Exposure $(k,n)$ we define the installments $R_{k,n}(t)$ as
\begin{equation} \label{eq:Rkn}
	R_{k,n}(t) = C_{k,n}(t) + I_{k,n}(t)
\end{equation}
where $C_{k,n}(t)$ and $I_{k,n}(t)$ are the quota capital and interest of the Exposure amortization schedule. We assume that $C_{k,n}$ and $I_{k,n}$ be zero outside the time interval where they are defined, i.e $C_{k,n}(t)=I_{k,n}(t)=0$ for $t<\hat{\tau}^*$ or $t>T_{k,n}$. Notice that with $\hat{\tau}^* = 0$, the first installment is due at the beginning of the subsequent period $t = 1 \dots T$. Then, the total capital due by Exposure $(k,n)$ is
\begin{equation} \label{eq:Ckn}
	C_{k,n} = \sum_{t=0}^{T} C_{k,n}(t)
\end{equation}
and the outstanding capital of Exposure $(k,n)$ is
\begin{equation} \label{eq:OCnk}
	OC_{k,n}(t)= C_{k,n} - \sum_{\tau=0}^t C_{k,n}(\tau)
	= \sum_{\tau=t+1}^T C_{k,n}(\tau)
\end{equation}
while the total securitization capital is
\begin{equation} \label{eq:C}
	C = \sum_{k=1}^{K} \sum_{n=1}^{N_k} C_{k,n}
\end{equation}
The total interest due by Exposure $(k,n)$ is
\begin{equation} \label{eq:Ikn}
	I_{k,n} = \sum_{t=0}^{T} I_{k,n}(t)
\end{equation}
and the outstanding interest of Exposure $(k,n)$ is
\begin{equation} \label{eq:OInk}
	OI_{k,n}(t)= I_{k,n} - \sum_{\tau=0}^t I_{k,n}(\tau)
	= \sum_{\tau=t+1}^T I_{k,n}(\tau)
\end{equation}
while the total securitization interest is  
\begin{equation} \label{eq:I}
	I = \sum_{k=1}^{K} \sum_{n=1}^{N_k} I_{k,n}
\end{equation}
We can then define the total securitization outstanding balance as  
\begin{equation} \label{eq:OBt}
	OB(t) = OC(t) + OI(t) = \sum_{k=1}^{K} \sum_{n=1}^{N_k} OC_{k,n}(t) + OI_{k,n}(t)
\end{equation}
We call a securitization \say{Islamic} when the total interest $I_{k,n}(t)$ of any Exposure ($N,K$) is always $0 \ \forall \ t$, and \say{European} otherwise. Independently if the securitization were to be Basic or Rolling, Islamic or European, the equations in this paper would hold true.

\subsection{Securitization: liability-side basic variables} \label{Subsec: C1 - Liability Inputs}
Let a securitization liability-side be composed of $P \ge 2$ Positions (see definition in Sec. \ref{Sec: Step 5}). Each of the $p=1,\dots, P$ Positions is amortized over $T_{p}$ months. All $P$ Positions start their amortization period at the same time $\hat{\tau}^*$, and their maximum duration is
\begin{equation} \label{eq:TP}
	TP = \max_p \, T_p
\end{equation} 
where $TP \in N$ and $TP \leq T$. Indeed, because there is a mismatch between the inbound and the outbound cash flow lifespan, often $TP < T$: for example, it might happen that the securitization Positions' lifespan be $15$ years while some Exposures might pay installments for $17$ years. Currently, most practitioners would discount the extra 2-years Exposures' cash flows from year $17$-th back to year $15$-th implicitly assuming that the securitization will be able to sell those extra cash flows to interested $3$-rd parties when the time come. This approach is not consistent with a pure cash-flow method, that requires to consider only cash flows that will certainly materialize within the Position maximum duration $TP$. Therefore, for the purpose of this paper, any inbound cash flow obtained after $TP$ will not be considered in computing the Positions' outbound cash flows (i.e. $OCF(t) = 0 \ \forall \ t > TP $). 

\subsection{Inputs, Events, Scenarios \& Clusters definition} \label{Subsec: C1 - IESC}
We define as a \textbf{Base Input} $\phi$ any initial parameter to compute the inbound cash flows defined at the time $\hat{\tau}^*$ of structuring. We call $\Phi$ the set of those unique Base Inputs. 
We define an \textbf{Event} $\lambda$ to be any action or situation that might affect, in positive or negative, the Base Inputs at a certain time $\hat{t}$. We call the set of those unique Events \textbf{List of Events} $\Lambda$ or \textbf{LE}. An Event of the LE ($\lambda \in \Lambda$) may affect each Exposure $(k,n)$ at a certain time $\hat{t}_{k,n}^{\lambda}$, where $0 \leq \hat{t}_{k,n}^{\lambda} \leq T_{k,n}$. Let us recall that the Kronecker delta symbol is defined as
\begin{equation} \label{eq:delta}
	\delta_{a,b}=
	\begin{cases}
		0 & \text{if $a \neq b$} \\
		1 & \text{if $a = b$}
	\end{cases}
\end{equation}
while the Heaviside theta symbol is defined as
\begin{equation} \label{eq:theta}
	\theta_{a,b}=
	\begin{cases}
		0 & \text{if $a < b$} \\
		1 & \text{if $a \geq b$}
	\end{cases}
\end{equation}
Thus, if any Event $\lambda$ of the List of Event $\Lambda$ affects the performance of the Exposure $(k,n)$ from (or at) happening at time $\hat{t}^\lambda_{k,n}$, then $\delta_{t,\hat{t}^\lambda_{k,n}}$ allows us to account for the impact of \emph{spot} Events, while $\theta_{t,\hat{t}^\lambda_{k,n}}$ allows us to account for \emph{continuous} Events whose impact extends from time $\hat{t}^\lambda_{k,n}$ on. In the following, to simplify the notation we indicate in formulas $\hat{t}_{k,n}^{\lambda}$ with $t_\lambda$ when it does not give rise to ambiguities.

We define a \textbf{Scenario} as a combination of the possible Events affecting the Base Inputs of all the Exposures $N$. We define the \textbf{Base Scenario} as the unique Scenario that is unaffected by any Event.  
The inbound cash flows computed in the Base Scenario can be considered as the initial estimated \say{budget} at the time $\hat{\tau}^*$ of structuring of the securitization. The inbound cash flows computed in any other Scenarios $s$ will diverge from the Base Scenario for a certain amount: if it is positive it is a gain while if it is negative it is a loss. In all other Scenarios, the probability of each Event affecting each Base Input $\phi$ of any Exposure $(k,n)$ differs according to their respective risk profile. In every Scenario, each Exposure $(k,n)$ can be affected over time by one or more elements of the List of Events. Due to the combinatorial explosion problem, it is impossible to use a brute force approach to enumerate the total number of Scenarios $|S|$ underlying the securitization (or even only  the number of Scenarios $|S_k|$ of portfolio $k$). Thus, it is necessary to use a sampling approach like the Monte Carlo Method. The reason is simple: if we consider only \textbf{IF} an element of the List of the Events materializes, the number of potential Scenarios $|S|$ depends on the number of elements $NE$; such quantity grows exponentially with the number $N_k$ of Exposures composing the $K$ portfolios as
\begin{equation} \label{eq:IFS}
	|S|= \prod_{k=1}^K NE^{N_k} 
\end{equation}

On the other hand, for the PEAL Method, it is not only relevant \textbf{IF} an Exposure is affected by an Event, but also \textbf{WHEN}: the impact on its performance can be sensibly different if an Event happens at time $t$ or at $t+12$. Therefore, the probability of the Scenario's cash flows depends on the probability of each Event impacting over time each Exposure $(k,n)$ composing that Scenario. Thus, the number of Scenarios $|S|$ for a securitization composed of $K$ portfolios, each with $N_k$ Exposures with the same amortization period $T_{k,1} = T_{k,2} \dots = T_{k}$ and the same number of Events $NE$ is 
\begin{equation} \label{eq:STS}
	|S|= \prod_{k=1}^K ((NE-1) \cdot T_{k} +1)^{N_k} 
\end{equation}
Thus, for example for a securitization composed of just one portfolio $K = 1$, with $N_k = 100$ Exposures, with an average $T_{k,n}$ between $36$ and $60$ months and a LE composed of only 3 elements ($\Lambda = 3$), the total number of Scenarios $|S|$ range between $73^{100}$ and $121^{100}$, that is well above the total estimated number of atoms in the universe (i.e.  $10^{82}$). Therefore, direct enumeration of all possible Scenarios is out of reach and it becomes mandatory to use an indirect approach for the computations, like the Monte Carlo Method. 

Finally, we define a \textbf{Cluster} as a set of homogeneous Exposures with the same risk profile: thus, we consider each portfolio $k$ to correspond to a single Cluster. For example, let a securitization originate from a single portfolio whose Exposures can be described by two different probability density functions (\textbf{PDF}): since the portfolio is described by two different Clusters, we consider them as two separate portfolios, and thus $K = 2$.   

\section{The PEAL Method to structure a securitization} \label{Sec: C2 - PEAL Method}
After having defined in Sec. \ref{Subsec: C1 - SecDef} the cases under which a transaction or a scheme must be considered a securitization, this Section introduces the PEAL Method structuring process. In abstract terms, a securitization has a balance sheet similar to the one of a traditional company: on the asset side there are the Exposures that generate inbound cash flows, and on the liability side there are the Positions to whom the securitization, at certain conditions, has the obligation to pay outbound cash flows. Such outbound payments must follow a univocal and unambiguous order, such that it must always be possible to describe it as a deterministic algorithm\footnote{Notice that a strict usage of a deterministic algorithm to describe payments' priority removes ambiguities on who must be paid first and avoids litigations.}. Equity is often negligible, thus we will not consider it in the modelling. 

To date, the established structuring practice for the asset side often uses a \emph{macro-approach} that entails that the total inbound cash flows be impacted by the negative Events historical loss averages. Thus, on the liability side, the documentation provided to investors - commonly known as \emph{investment memorandum} - only shows the average Scenario without any data related to any risk metric of the Positions, especially to time-dependent metrics, that would allow investors and supervisory authorities to be continuously updated on the evolution of the Positions' performances. Indeed, such metrics are nowadays almost impossible to calculate with the current \emph{macro-approach}. 

The PEAL Method, by considering on the asset side the statistical properties of every single Exposure $N_k$ in different Scenarios $S$, adopts a \emph{micro-approach} for calculating the securitization risk analysis on the liability side. This is analogous to the development of statistical mechanics, an area of physics that, by studying the behavior of large groups of microscopic components (atoms, molecules) has allowed to explain the laws of classical thermodynamics, which studies the relationships among macroscopic quantities characterizing materials like temperature, pressure, and heat capacity. It is worth noting that there is an exponential number of micro-states in statistical mechanics, which contributes to the complexity of the system. At the same time, such complexity enables better characterization of the fluctuations of the system. On the same footing, this paper considers the statistical dynamics of the exponential number of possible outcomes to better characterize the ongoing Positions' risk profiles and time Features.

\subsection{Structuring a securitization} \label{Subsec: C2 - Steps}
In the PEAL framework a \textbf{Structuring Method} is a set of mathematical rules to build a robust process where the total inbound cash flows are allocated as outbound cash flows among the different Positions respecting Reg 2402 provisions. All Structuring Methods follow the same 10-steps process. There are 4 Steps to characterize the asset side:
\begin{itemize}
	\item \textbf{Step 1}: select the Exposures Type;
	\item \textbf{Step 2}: select the LE $\Lambda$ and generate the Scenarios $\mathcal{S}$; \footnote{The set $\mathcal{S}$ must be a representative statistical sample of all the possible Scenarios $S$.}
	\item \textbf{Step 3}: compute the basic inbound building-blocks (BIB);
	\item \textbf{Step 4}: compute the composite inbound building-blocks (CIB).
\end{itemize}
Then there are 4 Steps to characterize the liability side:
\begin{itemize}
	\item \textbf{Step 5}: select the number of X Cost and Y Note Positions;
	\item \textbf{Step 6}: design the Positions to absorb the Tranches;
	\item \textbf{Step 7}: dimension the Gross Cost (GC) and Gross Notes (GN);
	\item \textbf{Step 8}: compute the Net Cost (NC) and Net Notes (NN).
\end{itemize}
Finally, there are 2 Steps to optimize the securitization Structuring Methods:
\begin{itemize}
	\item \textbf{Step 9}: compute the relevant Features;
	\item \textbf{Step 10}: optimize the chosen Features.
\end{itemize}
In the next 10 Sections, we describe each of the above steps more in detail. 

\section{Step 1: Select the Type of Exposures} \label{Sec: Step 1}
Let a group of Exposures that exhibit similar characteristics be defined as a \textbf{Type}. The 1st Step of any Structuring Method is to select, for each portfolio $k$, the Type of Exposures (\textbf{TE}). The following is a non-exhaustive list of the eleven most used Types in Italy to date:
\begin{enumerate}
	\item \textbf{Corporate Loan} (CL): loan provided to a company;
	\item \textbf{Mortgages Loan} (ML): loan with an immovable asset as collateral;
	\item \textbf{Auto Loan} (AL): loan with a car or truck as collateral;
	\item \textbf{Student Loan} (SL): loan provided to a student for education;
	\item \textbf{Credit Card Loan} (CC): money borrowed per credit card expenses;
	\item \textbf{Cessione Quinto Pensione} (QP): pension-backed loan;
	\item \textbf{Cessione Quinto Salario} (QS): salary-backed loan;
	\item \textbf{Real Estate} (RE): property-generating income (e.g. lands, buildings); 
	\item \textbf{Energy Estate} (EE): equipment producing clean energy;
	\item \textbf{Exotic Asset} (EA): perishable assets (e.g. wine, reggiano parmesan);
	\item \textbf{Negative Event} Exposures (NE): any loan that is affected by a negative Event by the time of structuring $\hat{\tau}_k$ (so-called \textit{unlikely-to-pay} or \textit{non-performing-exposures}).
\end{enumerate}
Notice that, although it is mathematically possible to structure a securitization composed of $K$ portfolios with \textit{heterogeneous} Types, European supervisory authorities have forbidden this provision in Reg 2402. Thus, to date all $K$ portfolios composing an European securitization must be \textit{homogeneous} in Type. Notice that Clusters further refine the definition of Types: in fact, the Exposures of the same Cluster not only have the same Type but also the same risk profile. Thus, a securitization based on Auto Loans with ratings A and B, would consist of $K=2$ portfolios (Clusters) with the same Type.

\section{Step 2: Select the LE \& generate the Scenarios} \label{Sec: Step 2}
The 2nd Step of any Structuring Method is to select, for each portfolio $k$, the List of Events ($LE$) $\Lambda$ that best describe the selected Type of Exposures, used to compute the Clusters. Some of the most relevant Events that might affect each Type of Exposure are listed in Table \ref{Tab: List of Events} in Appendix \ref{Appendix: List of Events}. Consider that any Structuring Method might use only a subset of the List of Events $\Lambda$ in Table \ref{Tab: List of Events}. As an example, if the List of Events were just: prepayments (\say{pe}) and defaults (\say{de}), then the $\Lambda=\{pe, de\}$. Once selected the $LE$, the arranger must generate the relative Scenarios via Monte Carlo, respecting the Clusters' risk profile distributions and correlations.

\section{Step 3: Basic inbound building-blocks (BIB)} \label{Sec: Step 3}
As will be further explained in this Section, the inbound cash flows can be allocated to 3 basic mutually exclusive and commonly exhaustive \emph{building-blocks} (the \textbf{Basic Inbound building-Blocks} or \textbf{BIB}) that allow designing uniquely the securitization asset-side: the assets (\textbf{A}), the losses (\textbf{L}); and the events recovery (\textbf{E}). The 3rd Step of any Structuring Method is to compute the BIB. See Sec. \ref{Subsec: Step 9 - Performance} for the definition of Full-Performing, Performing, Non-Performing, and Super-Performing Exposures.

\subsection{Asset building-block (A)} \label{Subsec: Step 3 - A}
Let the inbound cash flows of the $(K,N)$ Exposures in the Base Scenario be defined as Gross Asset $GA(t)$. Exploiting this definition, the Base Scenario $b$ inbound cash flows of an Exposure $(k,n)$ is
\begin{equation} \label{eq:Rbkn}
	R_{k,n}^{(b)}(t) = R_{k,n}(t) = C_{k,n}(t) + I_{k,n}(t)
\end{equation}
while the overall securitization Base Scenario inbound cash flows is
\begin{equation} \label{eq:gat}
	GA(t) = \sum_{k=1}^K \sum_{n=1}^{N_k} R_{k,n}^{(b)}(t)
\end{equation}
Then, the Scenario $s$ inbound cash flows of a Performing Exposure $(k,n)$ is
\begin{equation} \label{eq:Rskn ext}
	R_{k,n}^{(s)}(t) = C_{k,n}(t) \cdot \prod_{\lambda \in \Lambda}(1-\theta_{t,t^c_{\lambda,s}}) + I_{k,n}(t) \cdot \prod_{\lambda \in \Lambda}(1-\theta_{t,t^i_{\lambda,s}})
\end{equation}
while the overall securitization Scenario $s$ inbound cash flows is
\begin{equation} \label{eq:at}
	A^{(s)}(t) = \sum_{k=1}^K \sum_{n=1}^{N_k} R_{k,n}^{(s)}(t)
\end{equation}
that is the Asset building-block. Notice that $\lambda$ is any Event of the List of Events $\Lambda$, affecting in a particular Scenario the Exposure $(k,n)$; and $\theta_{t,t_\lambda}$ is the Heaviside function as explained in Sec. \ref{Subsec: C1 - IESC}. There exist Events that may impact only the capital $C_{k,n}(t)$ at time $t_\lambda^c$ or only the interest $I_{k,n}(t)$ at time $t_\lambda^i$. Considering that the $\prod_{\lambda \in \Lambda}(1-\theta_{t,t_\lambda})$ is the same as multiplying the cash flows for $(1-\theta_{t,\hat{t_\lambda}})$, where $\hat{t_\lambda}$ is the first occurrence in each Scenario $s$ of any Event. We can then simplify Equation \eqref{eq:Rskn ext} to
\begin{equation} \label{eq:Rskn}
	R_{k,n}^{(s)}(t) = C_{k,n}(t) \cdot (1-\theta_{t,\hat{t}^c_{\lambda,s}}) + I_{k,n}(t) \cdot (1-\theta_{t,\hat{t}^i_{\lambda,s}})
\end{equation}

\subsection{Loss building-block (L)} \label{Subsec: Step 3 - L}
Let the Scenario $s$ inbound cash flows of Non-Performing \& Super-Performing Exposures be defined as the Losses building-block $L^{(s)}(t)$. Exploiting this definition, the Scenario $s$ inbound cash flows of a Non-Performing or Super-Performing Exposure $(k,n)$ is
\begin{equation} \label{eq:Lskn}
	L_{k,n}^{(s)}(t) = R_{k,n}^{(b)}(t) - R_{k,n}^{(s)}(t)
\end{equation}
then Scenario $s$ total losses are
\begin{equation} \label{eq:lt}
	L^{(s)}(t) = \sum_{k=1}^K \sum_{n=1}^{N_k} L_{k,n}^{(s)}(t) = GA(t) - A^{(s)}(t)
\end{equation}
When in a particular Cluster all the ($\lambda \in \Lambda$) Events affecting the Exposures of the $K$ portfolios are MECE, the previous Equation \eqref{eq:lt} becomes
\begin{equation} \label{eq:lt mece}
	L^{(s)}(t) = \sum_{k=1}^K \sum_{n=1}^{N_k} R_{k,n}^{(b)}(t) \cdot \sum_{\lambda \in \Lambda} \theta_{t,t_{\lambda,s}} 
\end{equation}
where $\sum_{\lambda \in \Lambda} \theta_{t,t_\lambda} \in\{0,1\}$. Then, the Scenario cumulative Loss is
\begin{equation} \label{eq:lcum}
	L^{(s)} = \sum_{t=0}^T L^{(s)}(t)
\end{equation}
Exploiting $L^{(s)}$ definition is possible to establish a \emph{liability-independent} metric to compute the asset-side losses. Indeed, any Financial Institution, given the same assets inputs and with comparable risk models, would get an equivalent $L^{(s)}(t)$ distribution, despite substantially different Waterfall Configurations, and thus different payment priorities (see definition in Sec. \ref{Sec: Step 6 - Designing}).

\subsection{Event recovery building-block (E)} \label{Subsec: Step 3 - E}
Let the \emph{spot} inbound cash flows mitigating negative or deriving by positive Events in the Scenario $s$ be defined as the event recovery building-block $E^{(s)}(t)$.  For example, in Corporate Loans Type if the List of Events where $\Lambda = \{ pe, de, eu \}$, then the Exposure $(k,n)$ event recovery $E_{k,n}(t)$ would be the sum of \say{spot} recoveries such as the prepayments $p_{k,n}(t)$ and the recovery of collateral $d_{k,n}(t)$ as
\begin{equation} \label{eq: eskn}
	E^{(s)}_{k,n}(t) = p_{k,n}^{(s)}(t) + d_{k,n}^{(s)}(t) 
\end{equation}
while the total inbound event recovery in Scenario $s$ would be
\begin{equation} \label{eq: es}
	E^{(s)}(t) = \sum_{k=1}^K \sum_{n=1}^{N_k} E^{(s)}_{k,n}
\end{equation}

\subsection{Disambiguating spot vs non-spot cash flows} \label{Subsec: Step 3 - Disambiguating BIB}
In the previous Sections we have explained that \emph{non-spot}\footnote{Non-Spot: cash flows different from zero more than once in different time $t$.}  inbound cash flows shall be described by the Gross Assets $GA(t)$, the Assets $A(t)$ and the Losses $L(t)$. On the other hand, the \emph{spot}\footnote{Spot: cash flows different from zero just at one specific time $t$.} inbound cash flows shall be described by the event recovery $E(t)$.     

For example, if we consider the Corporate Loans Type of Exposures we know from Table \ref{Tab: List of Events} that the main Events that apply are $\Lambda = \{ pe, de, trl, eu \}$\footnote{Events: prepayments, defaults, return to life, and variation of the euribor}. Thus, an Exposure $(k,n)$ is expected to prepay voluntarily at time $t_{k,n}^{pp}$ if it is affected by the prepayment Event $pe$: whilst the negative effect impacts the Assets $A(t)$, the positive prepayment cash flow is \emph{spot} at time $t_{k,n}^{pp}$ and thus shall be absorbed by the event recovery building-block $E(t)$ as $p(t)$. The same approach can be used for the default Event $de$, so the positive recovery of collateral cash flow is \emph{spot} at time $t_{k,n}^{rec}$ and thus shall be absorbed by the event recovery building-block $E(t)$ as $d(t)$. On the other hand, if an Exposure $(k,n)$ that has defaulted at time $t_{k,n}^{df}$, is affected by the \say{return-to-life} Event $rtl$ at a time $t_{k,n}^{rtl}$ such that $t_{k,n}^{df} \leq t_{k,n}^{rtl} < t_{k,n}^{rec}$, then $d(t) = 0$ and $A(t)$ will have the form:
\begin{equation} \label{eq: rtl}
	R_{k,n}^{(s)}(t) = R_{k,n}(t) \cdot (1-\theta_{t,t_{k,n}^{df}} + \theta_{t,t_{k,n}^{rtl}} )
\end{equation}
where $\theta_{t,t_{k,n}^{rtl}}$ \say{reactivates} the cash flows after the recovery time $t_{k,n}^{rtl}$. 

The same dynamic applies to the Euribor Event $eu$, where we expect the Euribor to change overtime versus its Base Input, and thus we need to account for the change of installment for each Exposure $(k,n)$ whose interest is dependant on the change of euribor. A positive or negative variation of the euribor from its Base Input value/function changes the Exposure $(k,n)$ total installment and therefore we can consider that the impact of the Euribor Event $eu$ is \textit{non-spot} and thus shall be accounted in $A(t)$. Notice that if the new $A(t)$ is lower than the Base Input $GA(t)$, the $L_{k,n}(t \geq t_{k,n}^{eu} ) > 0$. On the other hand, if the new $A(t)$ is higher than the Base Input $GA(t)$, the $L_{k,n}(t \geq t_{k,n}^{eu} ) < 0$. Lastly, it is important to remember that a change in euribor shall not be immediately reflected in a change in the payments toward Positions: indeed, an increase in euribor is likely to increase the probability of default, and a decrease in euribor is likely to trigger an increase of prepayments, with respect to the default and the prepayment distribution values that were expected initially. Further analyses on the implication of the Euribor Event against the excess losses deriving from changes versus the Base Inputs shall be done.

The disambiguation logic applied in the previous example for Corporate Loans can be applied to any Event in any Type of Exposure. When deciding whether an inbound cash flow shall be allocated to Gross Assets $GA(t)$, Assets $A(t)$ or Events recovery $E(t)$, always consider if its cash flows are indeed \emph{spot} or \emph{non-spot}, and allocate accordingly the relative cash flows to the proper inbound building-block.

\section{Step 4: Composite inbound building-blocks (CIB)} \label{Sec: Step 4}
As will be further explained in this Section, the basic inbound building-blocks can be used to compute different composite inbound building-blocks (CIBs). The 4th Step of any Structuring Method is to compute all the following CIB:
\begin{enumerate}
	\item $ICF(t)$: total inbound cash flows as in Eq. \eqref{eq: ICFt};
	\item $TAF^{(s)}(t)$: total available funds as in  Eq. \eqref{eq:taft};
	\item $TL^{(s)}(t)$: inbound total loss as in Eq. \eqref{eq: TLt};
	\item $TNL^{(s)}(t)$: inbound total net loss as in Eq. \eqref{eq: TNLt};
	\item $Tranches$: first loss, second loss, and complementary loss tranches.
\end{enumerate}

\subsection{Optimization inbound building-block} \label{Subsec: Step 4 - OIB}
Notice that, in calculating all the following composite inbound building-blocks, we will introduce a new block that is dimensioned at the end of the structuring process, in the optimization phase of Step 10 (see Sec. \ref{Sec: Step 10}). This \say{new} block represents an inbound cash flow paid by a 3rd-party, often \emph{una-tantum} at time $t=0$, as a lump sum that is neither reimbursed nor remunerated (e.g. the upfront payment provided to cover the first agents' expenses in Negative Event Exposures) and will be called $zero(t)$ or $z(t)$. In the first structuring cycle, such amount shall be considered null ($z(t) = 0$), and only in the optimization phase the arranger will have the choice to dimension $z(t)$. To date, we define the Base Scenario $b$ initial endowment inbound cash flows as $z^{(b)}(t)$ and the Scenario $s$ initial endowment inbound cash flows $z^{(s)}(t)$. Normally, the difference between these two amount is always zero, but there might exist theoretical situations where ($z^{(b)}(t) > 0$) and ($z^{(s)}(t) = 0$) due to unexpected situations like the bankruptcy of the trustee hosting the cash flows or the default of the bonds in which the cash reserve is invested. In some cases, the difference might be due to a delay in the upfront payment by the 3-rd party.

\subsection{Total Inbound Cash Flows (ICF)} \label{Subsec: Step 4 - ICF}
We define the Base Scenario total inbound cash flows $ICF(t)$ as  
\begin{equation} \label{eq: ICFt_ext}
	\begin{split}
		ICF(t) & = \; GA(t) + E^{(b)}(t) + z^{(b)}(t) \\
	\end{split}
\end{equation}
considering that in the Base Scenario $E^{(b)}(t) = 0 \ \forall t$ then 
\begin{equation} \label{eq: ICFt}
	ICF(t) = GA(t) + z^{(b)}(t) 
\end{equation}
that represents the maximum initial endowment given the Base Inputs.

\subsection{Total Available Funds (TAF)} \label{Subsec: Step 4 - TAF}
We define the inbound cash flows that can be allocated at any time $t$ as an outbound cash flows to the Positions in any Scenario $s$ as total available funds $TAF^{(s)}(t)$ as
\begin{equation} \label{eq:taft}
	\begin{split}
		TAF^{(s)}(t) & = \; A^{(s)}(t) + E^{(s)}(t) + z^{(s)}(t) \\
	\end{split}
\end{equation}

\subsection{Total Loss (TL)} \label{Subsec: Step 4 - TL}
We define the sum of Scenario $s$ Losses $L^{(s)}(t)$ and the difference between the Base Scenario $b$ and the Scenario $s$ optimization inbound block $z(t)$ plus the Super Senior Embedded Position $SSE^{(s)}(t)$ (see definition in Sec. \ref{Subsec: Step 5 - Embedded Positions}) as Total Loss $TL^{(s)}(t)$ 
\begin{equation} \label{eq: TLt}
	\begin{split}
		TL^{(s)}(t) & = \; L^{(s)}(t) + [z^{(b)}(t) - z^{(s)}(t)] + SSE^{(s)}(t) \\
		& = \; [GA(t) - A^{(s)}(t)] + [z^{(b)}(t) - z^{(s)}(t)] + SSE^{(s)}(t)\\
		& = \; [GA(t) + z^{(b)}(t)] - [A^{(s)}(t) + z^{(s)}(t)] + SSE^{(s)}(t)\\
		& = \; ICF(t) - [TAF^{(s)}(t) - E^{(s)}(t)] + SSE^{(s)}(t)\\
		& = \; [ICF(t) - TAF^{(s)}(t)] + E^{(s)}(t) + SSE^{(s)}(t)\\
		& = \; TNL^{(s)}(t) + E^{(s)}(t) + SSE^{(s)}(t)\\
	\end{split}
\end{equation}
where the $ICF(t)$ is Equation \eqref{eq: ICFt}, the $TAF^{(s)}(t)$ is Equation \eqref{eq:taft}, the $E^{(s)}(t)$ is Equation \eqref{eq: es}, the $TNL^{(s)}(t)$ is the \emph{Total Net Loss} of Equation \eqref{eq: TNLt}, and $SSE^{(s)}(t)$ is Equation \eqref{eq: SSE}. 

Notice that the Super Senior Embedded Position $SSE^{(s)}(t)$ components are either \textit{additional expenses} (see Excessive Costs or Cost of Recovery in Sec. \ref{Subsec: Step 5 - Embedded Positions}) or \textit{reduced outbound cash flows} for the Positions (see Excess Recovery in Section \ref{Subsec: Step 5 - Embedded Positions}) that reduce the total available funds $TAF^{(s)}(t)$. Therefore, $SSE^{(s)}(t)$ shall be treated as \say{additional losses} and shall be added to the total loss $TL^{(s)}(t)$ to provide a clear picture of the securitization ongoing riskiness to the institutional investors and the public authorities. Finally, it is worth noticing that, by default, the Losses $L^{(s)}(t)$ are always \say{smooth} whilst the Total Losses $TL^{(s)}(t)$, absorbing the \emph{spot} inbound cash flows, can have loss spikes. As shown in Sec. \ref{Subsec: Step 4 - Tranching}, by computing the Tranching on the distribution both of the $L^{(s)}(t)$ and of the $TL^{(s)}(t)$ it is possible to understand what percentage of risk depends on the \emph{non-spot} functions and what percentage from the \emph{spot} functions. 

\subsection{Total Net Loss (TNL)} \label{Subsec: Step 4 - TNL}
We define the difference between the Scenario $s$ losses $L^{(s)}(t)$ and the Scenario $s$ events recovery $E^{(s)}(t)$  as the net loss $NL^{(s)}(t)$
\begin{equation} \label{eq: NLt}
	NL^{(s)}(t) = L^{(s)}(t) - E^{(s)}(t) 
\end{equation}
Then, we define the difference between the Scenario $s$ total losses $TL^{(s)}(t)$ and the Scenario $s$ events recovery $E^{(s)}(t)$ as the total net loss $TNL^{(s)}(t)$
\begin{equation} \label{eq: TNLt}
	\begin{split}
		TNL^{(s)}(t) & = \; TL^{(s)}(t) - E^{(s)}(t) \\     
		& = \; [L^{(s)}(t) + z^{(b)}(t) - z^{(s)}(t) + SSE^{(s)}(t)] - E^{(s)}(t) \\
		& = \; [GA(t) - A^{(s)}(t)] + [z^{(b)}(t) - z^{(s)}(t)] - E^{(s)}(t) + SSE^{(s)}(t)\\
		& = \; [GA(t) + z^{(b)}(t)] - [A^{(s)}(t) + E^{(s)}(t) + z^{(s)}(t)] + SSE^{(s)}(t)\\
		& = \; ICF(t) - TAF^{(s)}(t) + SSE^{(s)}(t)\\
	\end{split}
\end{equation}
When relevant, prefer the total loss $TL^{(s)}(t)$ over the total net loss $TNL^{(s)}(t)$ to compute the Features, as a more robust metric to compare risk profiles across securitizations. 

\subsection{Total Loss Tranching} \label{Subsec: Step 4 - Tranching}
This Section is divided in 2 parts: in Sec. \ref{Subsubsec: Step 4 - Tranching Definition} we define the concept of first, second and complementary loss tranches from a regulatory perspective; and in Sec. \ref{Subsubsec: Step 4 - Substantial Margin} we explain how to compute the significant margin within the First Loss Tranche.

\subsubsection{Tranching definition} \label{Subsubsec: Step 4 - Tranching Definition}
As previously stated, Reg 2402, Article 2(1)(b), explains that in a securitization the subordination of tranches determines the distribution of losses during the \textit{ongoing life} of the transaction scheme. From this article we can infer two things: (i) the risk is segmented in more than one tranche; (ii) the tranches shall be evaluated on an \say{ongoing basis}. In this paper, we will consider 3 types of tranches: the first loss tranche (\textbf{FLT}), the second loss tranche (\textbf{SLT}), and the complementary loss tranche (\textbf{CLT}).

Reg 2402, Article 2(18), defines the \textit{first loss tranche} as \say{the most subordinated tranche in a securitization that is the first tranche to bear losses incurred on the exposures and thereby provides protection to the second loss and, where relevant, higher ranking tranches}. What we can deduce from Article 2(18) is that in a securitization there must be at least two Positions ($P \geq 2$) absorbing the different tranches. Therefore, each tranche must be covered by at least one Position, and each Position can \say{absorb} risk from more than one tranche. In addition, European Capital Requirement Regulation No 575/2013 consolidated version of January 2024 (\textbf{CRR}), Article 244(1)(a), explains that an \say{originator institution of a traditional securitization may exclude underlying exposures from its calculation of risk-weighted exposure amounts if [$\dots$] significant credit risk associated with the underlying exposures has been transferred to third parties}. Article 244(2)(b) explains that \say{significant credit risk shall be considered as transferred} if \say{the originator can demonstrate that the exposure value of the first loss tranche exceeds a reasoned estimate of the expected loss of the underlying exposure by a substantial margin}. Thus the CRR links the $FLT$ to the Exposure expected loss (\textbf{EL} or $\mu$) plus a substantial margin ($sm$). The CRR does neither provide further details on which distribution to compute the $EL$ nor on how to compute the \say{substantial margin}. For the purpose of this paper, as distribution we will use the total net loss $TNL^{(s)}(t)$, whilst for how to asses weather or not there exist $sm$ we recommend to see next Section (\ref{Subsubsec: Step 4 - Substantial Margin}). Putting aside for a moment the \say{significant margin}, the computation of the $FLT(t)$ is as follows
\begin{equation} \label{eq: FLT}
	FLT(t) = \max\left(0,\mu(t)\right)
\end{equation}
where
\begin{equation} \label{eq: ELt}
	\mu(t) = \left\langle TNL^{(s)}(t)\right\rangle
\end{equation}
where $\mu(t)$ is the mean with respect to the total net losses $TNL^{(s)}(t)$.

Whilst for the $FLT(t)$ there is an indication on how to compute it, for what concerns the second loss tranche $SLT(t)$, to the best of our knowledge,  there is no jurisprudence. After having interviewed tens of practitioners, we believe that the most used method to compute the SLT -- a measure of the Unexpected Loss -- is the \textit{expected shortfall}\footnote{The \say{expected shortfall} is also known as \say{conditional value at risk} \citep{Rockafellar2000}, or \say{super quantile} \citep{rockafellar2010buffered}.} or \say{value at risk} (\textbf{VaR}). Thus the SLT is the mean of the total net losses $TNL^{(s)}(t)$, restricted to the values exceeding a certain $\alpha \%$ quantile, minus the first loss tranche $FLT(t)$ as
\begin{equation} \label{eq: SLT}
	SLT(t) = VaR_{\alpha}(t)-FLT(t)
\end{equation}
where
\begin{equation} \label{eq: VAR}
	VaR_{\alpha}(t) = \mu(TNL^{(s)}(t) \ge \hat{q}_{\alpha})
\end{equation}
is the value at risk $VAR$ and 
\begin{equation} \label{eq: qhat}
	\hat{q}_{\alpha} = F_{TNL}^{-1}(\alpha)
\end{equation}
is $TNL^{(s)}(t)$ cumulative distribution function at $\alpha \%$ quantile. 

Finally, we define the $CLT(t)$ as the difference between the total inbound cash flows $ICF(t)$ and the other tranches as  
\begin{equation} \label{eq: CLT}
	CLT(t) = ICF(t)-SLT(t)-FLT(t)
\end{equation}
Notice that the $CLT(t)$ covers extreme losses like the one that materializes in the theoretical scenario where all Exposures turn Non-Performing at $t = 0$ and all $E^{(s)}(t) = 0 \ \forall t$. In this theoretical Scenario, as Equation \eqref{eq: TNLt} shows, being the $A^{(s)}(t) = E^{(s)}(t) = 0 \ \forall t$ then the total net loss is equal to the inbound cash flows $TNL(t) = ICF(t)$. The complementary loss tranche is normally the tranche that is covered by the very first Positions, like the costs and the \say{first} note Positions (see next Sec. \ref{Sec: Step 5} for further details). 

In any Structuring Method, for the sake of transparency, consistency, and cross-comparability, it should become customary to define \textit{always} the $FLT$, the $SLT$, and the $CLT$ (together the \textbf{Tranches}). Indeed, these 3 categories exist in any Structuring Method since they are necessary to compute the regulatory capital (see Sec. \ref{Subsec: Step 5 - Positions Qualities} and Sec. \ref{Subsec: Step 9 - Regulatory Capital}). Considering that each Position (with $P \ge 2$) may absorb less or more than one of the above-mentioned 3 Tranches, the arranger must be transparent in indicating what percentage of each Tranche each Position absorbs. 

\subsubsection{FLT substantial margin} \label{Subsubsec: Step 4 - Substantial Margin}
In the previous Sec. \ref{Subsubsec: Step 4 - Tranching Definition} we explained that the CRR links the $FLT$ computation to the concept of a \say{significant margin} $sm$, without mathematically defining it. To the best of our knowledge, there is no conclusive evidence on how the Supervisory Authorities apply this concept, and yet it is a core aspect in computing regulatory capital (see Sec. \ref{Subsec: Step 9 - Regulatory Capital}). When the CRR entered into force in 2013, practitioners computed the $FLT$ on the cumulative PDF of the total net loss $TNL$ at the Positions' maximum duration $TP$
\begin{equation} \label{eq: TFLT}
	TFLT = \left\langle \sum_{t=0}^{TP} TNL^{(s)}(t)\right\rangle
\end{equation}
instead of on an ongoing basis as currently required by the Reg 2402, that entered into force in 2017. Thus, we propose a solution in implementing the \say{significant margin} of the 2013's CRR that takes into account the ongoing nature of the FLT defined by Reg 2402. Indeed, the following ratio
\begin{equation} \label{eq: sm}
	sm = \frac{\sum_{t=0}^{TP} FLT(t)}{TFLT}-1
\end{equation}
is $\geq 0$  when $FLT(t)$ is defined as equation \eqref{eq: FLT}, and $TFLT$ is defined as equation \eqref{eq: TFLT}. A criterion to understand if the margin is \say{significant} could be to check that it covers more than a standard deviation of the total first loss tranche (i.e. $sm \geq \sigma_{TFLT}/TFLT$). By definition, our $FLT$ of equation \eqref{eq: FLT} already has a \say{significant margin} embedded within its formulation versus the methodology applied back in 2013 when this concept was introduced. This is also why in the previous Sec. \ref{Subsubsec: Step 4 - Tranching Definition} we ignore the $sm$ parameter. 

\section{Step 5: Select the number of Positions} \label{Sec: Step 5}
As further explained in this Section, the total inbound cash flows $ICF(t)$ are allocated to one macro building-block called \say{Position}, composed of 2 regulatory building-blocks called \say{Cost} and \say{Note} Positions and 1 \say{ephemeral} called \say{Embedded} Position. The 5th Step of any Structuring Method is to select the number of $X$ Cost Positions and the number of $Y$ Note Positions. This Section is divided in 4 parts: in Sec. \ref{Subsec: Step 5 - Positions Definition} we define the concept of Position from a regulatory perspective; in Sec. \ref{Subsec: Step 5 - Positions Qualities} we define the concept of Position \say{Quality}; in Sec. \ref{Subsec: Step 5 - Embedded Positions} we define the concept of Embedded Positions; and in Sec. \ref{Subsec: Step 5 - Positions Computation} we explain how to compute the number of Positions.

\subsection{Position definition} \label{Subsec: Step 5 - Positions Definition}
Reg 2402 does not define the concept of \textit{positions}, instead its Article 2(19) only defines the concept of \textit{securitization position} as \say{an exposure to a securitization}. In the whole Reg 2402, the concept of \textit{position} is used equivalently and indistinctly to \textit{securitization positions}, thus we use the same hermeneutic interpretation of equating \textit{position} with \textit{securitization position}. 

European Regulation No. 2401/2017 (\textbf{Reg 2401}), Article 242(6) defines \textit{senior securitization position} as a Position \say{backed or secured by a first claim on the whole of the underlying exposures}, \textit{disregarding} for these purposes amounts due under interest rate or currency derivative contracts, \textit{fees} or \textit{other similar payments}, and irrespective of any difference in maturity with one or more other senior tranches with which that position shares losses on a pro-rata basis. The previous Article 242(6) points out something that is often overseen: \say{fees or other similar payments} has to be disregarded when assessing if a Position is \say{senior} or not. Indeed, in current Structuring Methods, servicing fees are often the ones with the utmost priority in the payments versus the most senior Note Positions. Thus, in the first place, if the servicing fees were not actually considered Positions by Reg 2401, then Article 242(6) would have not needed to carve them out from the \say{senior position} definition. It follows that Reg 2401 considers servicing fees a Position (\textbf{Cost Position}) not differently than the notes (\textbf{Note Position}).
Indeed, since servicing fees receive a fraction of the total inbound cash flows $ICF(t)$, they must be considered absorbing as well a fraction of credit risk (i.e. if the whole portfolio were to default at time $t = 0$, then also the servicing fees would not be paid accordingly to the Base Scenario). 

On the other hand, Reg 2401, Article 242(6), defines a \textit{mezzanine securitization position} as a \say{position [$\dots$] which is subordinated to the senior [$\dots$] and more senior than the first loss tranche}. Finally, neither Reg 2401 nor Reg 2402 defines the concept of \textit{junior securitization position} but generally refers to the First Loss Tranche. Then, we can \textit{assume} that a Position is \say{junior} if it absorbs the First Loss Tranche $FLT(t)$. Consequently, we can assume that a Position is \say{mezzanine} if it absorbs the Second Loss Tranche $SLT(t)$ and \say{senior} if it absorbs the Complementary Loss Tranche $CLT(t)$.

Notice that some practitioners do not consider the Costs as Positions because they net the total inbound cash flows of the various expenditures toward the servicers and agents before redistributing the cash flows to the Notes Positions. This practice goes against many International Financial Reporting Standards (IFRS) principles described into the \citep{IFRSFramework}, in particular the one that requires to \say{separately reporting the components of financial statements}. This principle emphasizes the importance of presenting revenues, expenses, and cash flows separately in financial statements rather than netting them together. It ensures transparency and provides users of financial statements with a clear understanding of an entity's financial performance and financial position. Therefore, the investment memorandum shall always respect the same IFRS principles and thus clearly shows the Cost Positions in at least 2 different forms: (i) the $x$-th Net Dimensioned Cost Positions $NC^{(s)}_x(t)$ as defined in Equation \eqref{eq: NC & NN}; and (ii) the ratio of $NC^{(s)}_x(t)$ to the total inbound cash flows $ICF(t)$. These 2 Features should be reported not only as their average value, but also as their empirical (i.e. from the simulated scenarios) ongoing probability density function, because in certain Scenarios the cost of recovery $CR^{(s)}(t)$ of defaulted Exposures might be so relevant to impact both the mezzanine and the senior Positions, especially in Negative Event Exposures. Thus, it is crucial to not only consider the Costs as Positions, but to compute all the above Features to enable the investors to make an informed judgement on the real intrinsic riskiness of the Notes they are purchasing.  

\subsection{Position Qualities} \label{Subsec: Step 5 - Positions Qualities}
Why is it important to define if a Position is \say{senior}, \say{mezzanine}, or \say{junior}? From a Structuring Method perspective, these regulatory definitions are mathematically irrelevant. Indeed, if at any time the Regulator were to decide to abolish these concepts, the securitization modelling would not be impacted. Therefore, the \say{seniority} is not an intrinsic element of any securitization, but more like a \emph{quality} that a Position obtains depending on multiple factors. Assigning the right \emph{quality} to a Position is not just an aesthetic quirk: in fact, to compute the Regulatory Capital (\textbf{RC}) we need to assess, for each Position, the relative \emph{risk weights} (\textbf{RWs}) applicable to a slice or to the Position as a whole. Indeed, as it can be seen in the table of Reg 2401 Article 259(1), the regulator has assigned different RWs to \emph{senior} or \emph{non-senior} Positions and thus assigning the right quality has direct implications on the RC computation. In conclusion, the \textit{quality} is not an embedded characteristic that each Position has by default, but depends on its Designing (i.e. defining the relative \say{location} of the X Cost relative to the Y Notes Positions, see next Sec. \ref{Sec: Step 6 - Designing}) and its Dimensioning (i.e. assigning a \say{value} to the Positions, see next Sec. \ref{Sec: Step 7 - Gross Dimensioning}) whose effect is only relevant when computing the RC. Given the same Dimensioning, a diverse Designing would assign to each Position a different \textit{quality}. 

See the next Sec. \ref{Subsec: Step 9 - Regulatory Capital} for a deep-dive on Regulatory Capital: for now, just remember that for the purpose of this paper, to each Position, in part or as a whole, is assigned  over time one (or more!) of the following 3 \textbf{Qualities}:
\begin{itemize} \label{Quality Description}
	\item \textbf{Senior}: if a Position slice absorbs a \% of $CLT(t)$;
	\item \textbf{Mezzanine}: if a Position slice absorbs a \% of $SLT(t)$;
	\item \textbf{Junior}: if a Position slice absorbs a \% of $FLT(t)$.
\end{itemize}
since, as it will be further explained in the next Sec. \ref{Sec: Step 6 - Designing}, each Position might absorb different tranches of the risk depending on their Designing. Therefore a Position might not receive the \say{quality} of Senior, Mezzanine or Junior as a whole, but each different \say{slice} absorbing a tranche of the FLT, SLT or CLT should receive the consequent \say{quality}.

Now, our statement about the Qualities might seem to contradict the previous Article 242(6) of Reg 2401, where it is clearly stated that a Cost Position cannot be considered Senior, whilst with our definition a Cost Position could get the quality of Senior if it were absorbing a slice of the complementary loss tranche $CLT(t)$. This contradiction will be resolved in the next Sec. \ref{Subsec: Step 9 - Regulatory Capital}: indeed, the Regulator removed Cost Positions \textit{ex-ante} from the computation of $RC$, whilst instead of making Cost Positions lose their Qualities \textit{ex-ante}, in the PEAL Method we assign to the Cost Positions a risk weight $RW = 0$ \textit{ex-post} as shown in Equation \eqref{eq: rcny}. In fact, as we will show in the next Sec. \ref{Subsec: Step 6 - Designing Definition}, our Structuring Method allows to design \say{vertical} Cost and Note Positions, and thus it becomes crucial to provide the securitization stakeholders with a transparent picture of which part of which Position has which Quality, irrespectively if a Position bears or not regulatory capital ($RC$). Indeed,  Qualities are currently used to intuitively understand the liability-side payment priorities of the Positions but, since our approach allows for \say{Vertical} Positions that can even be Senior, Mezzanine and Junior at the same time, the direct link between Qualities and payment priorities breaks. Thus, in general, Qualities should not be used as a proxy for the payment priorities, but only to calculate the Regulatory Capital.

\subsection{Embedded Positions} \label{Subsec: Step 5 - Embedded Positions}
Let a Position be \say{Embedded} if it is equal to zero for any $t$ in the Base Scenario: $EP^{(b)}(t) = 0 \ \forall \ t $. Embedded Positions can be different from $0$ only in some Scenario $s$: $EP^{(s)}(t) \geq 0$. Embedded Positions are divided into two: (i) Super Senior (\textbf{SSE}) and (ii) Super Junior (\textbf{SJE}). We introduced the Embedded Positions because currently they are not properly described within the documents provided to the stakeholders. Therefore, the information about these variables is often overseen and it becomes incredibly complex for 3rd-parties to clearly understand their impact on their overall returns. Following the same principles of clarity and transparency described in the previous Sec. \ref{Subsec: Step 5 - Positions Definition}, we believe that the introduction of the Embedded Positions makes the risk assessment in the different Scenarios $s$ way simpler for all stakeholders.

\subsubsection{Super Senior Embedded Positions} \label{Subsubsec: Step 5 - Super Senior Embedded Positions}
The Super Senior Embedded (\textbf{SSE}) are those Positions that have always the first claim on the total available funds $TAF^{(s)}(t)$, irrespectively from any possible Designing (see next Sec. \ref{Sec: Step 6 - Designing}). To date, we have identified only $3$ types of SSE that are the following:
\begin{itemize}
	\item \textbf{EC}: stands for Excessive Costs, which are all the \textit{una-tantum} fines issued by any authority in case of infringement of one or more national or sovra-national laws (e.g. tax fines) plus the legal costs to defend from such accusations. It is Super Senior because any legal fees have always the utmost priority in any possible Designing;
	\item \textbf{CR}: stands for Cost of Recovery, that are the costs that the securitization pays on an ongoing basis to recover the defaulted Exposures. It is Super Senior because the agents that recovery the distressed credits shall be paid before anyone else; 
	\item \textbf{ER}: stands for Excessive Recovery, which are the eventual excess cash flows that in each Scenario $s$ derives from the difference between an Exposure $(k,n)$ outstanding capital at the time of the default $OC_{k,n}^{(s)}(t^{df})$, plus the cumulative recovery Cost $CR_{k,n}^{(s)}(t)$, and the recovered value at the time of recovery $RV_{k,n}^{(s)}(t^{tr})$ as
	\begin{equation} \label{eq: ERt}
		ER^{(s)}_{k,n}(t) =  \max \left( 0, RV^{(s)}_{k,n}(t^{tr}) - OC^{(s)}_{k,n}(t^{df}) - \sum_{t=0}^{t^{tr}} CR_{k,n}^{(s)}(t) \right) \\
	\end{equation}
	It is Super Senior because the Excessive Recovery belongs to the respective Exposures from which they have derived and must be revert as soon as the excessive cash is made available to the securitization, without further ado.
\end{itemize}
Therefore, the Scenario $s$ Super Senior Embedded Position $SSE^{(s)}(t)$ is
\begin{equation} \label{eq: SSE}
	SSE^{(s)}(t) = EC^{(s)}(t) + \sum_{k = 1}^{K} \sum_{n = 1}^{N} CR_{k,n}^{(s)}(t) + ER_{k,n}^{(s)}(t) 
\end{equation}
whilst the average Scenario Super Senior Embedded Position $\overline{SSE}(t)$ is
\begin{equation} \label{eq: SSEm}
	\overline{SSE}(t) = \left\langle SSE^{(s)}(t) \right\rangle 
\end{equation}

\subsubsection{Super Junior Embedded Positions} \label{Subsubsec: Step 5 - Super Junior Embedded Positions}
The Super Junior Embedded (\textbf{SJE}) are those Positions that have always the last claim on the total available funds $TAF^{(s)}(t)$, irrespectively from any possible Designing (see next Sec. \ref{Sec: Step 6 - Designing}). To date, we have identified only $1$ type of SJE that is:
\begin{itemize}
	\item \textbf{B}: stands for Buffer, which are those excess cash flows that in each Scenario $s$ derive from the difference between $A^{(s)}(t)$ and $GA(t)$ as
	\begin{equation} \label{eq: EEt}
		B^{(s)}(t) =  \max \left( 0, A^{(s)}(t) - GA(t) \right) \\
	\end{equation}
	The $B^{(s)}(t)$ can be caused by an excess of euribor or other factors that generate a negative losses. The Buffer is Super Junior because it is an excess cash used to absorb the first loss tranche $FLT(t)$, even before any $p$-th Position, and it is normally paid out \textit{pari-passu} to the \say{lower-tier} Positions. 
\end{itemize}

\subsection{Compute the number of Positions} \label{Subsec: Step 5 - Positions Computation}
As detailed in the previous Sec. \ref{Subsec: Step 5 - Positions Definition}, there are of 2 types of Positions: those owned by the securitization servicers (\textbf{Cost Positions}) whose total number is described by the letter $X$, and those issued as notes and detained by bondholders (\textbf{Note Positions}) whose total number is described by the letter $Y$. Thus, the total number of Positions $P$ is
\begin{equation} \label{eq:position check}
	P = X + Y
\end{equation} 
Pursuant Reg 2402, a transaction is a securitization if and only if $P \ge 2$, of which one must be a Cost Position $X \ge 1$ and one a Note Position $Y \ge 1$. Indeed, securitization vehicles are empty shells and need servicers to operate them. Notice that, for the purpose of this paper, if there are $3$ service providers paid pari-passu, we can sum their cash flows and consider them being one single Cost Position when computing the payment priority waterfall. The same reasoning shall be applied to the notes: if two or more Note Positions are completely \say{pari-passu}, sharing the same risk and being paid with the same frequency, then they shall be deemed as one single Note Position, summing their cash flows.      

\section{Step 6: Positions' Designing to absorb the Tranches} \label{Sec: Step 6 - Designing}
Positions' Designing determines the way the inbound risk is distributed among the outbound Positions. This Section is divided in seven parts: in Sec. \ref{Subsec: Step 6 - Designing Definition} we define the process to organize the Positions to design a Waterfall Configuration; in Sec. \ref{Subsec: Step 6 - Virtual Positions} we introduce the concept of Virtual Positions; in Sec. \ref{Subsec: Step 6 - Horizontal Components} we introduce the concept of Horizontal Components; in Sec. \ref{Subsec: Step 6 - Vertical Components} we introduce the concept of Vertical Components; in Sec. \ref{Subsec: Step 6 - Positions' Designing} we introduce the concept of Positions' Designing; in Sec. \ref{Subsec: Step 6 - Explanation of the example} we analize in detail the illustrative Design of Figure \ref{fig: step 6}; and finally in Sec. \ref{Subsec: Step 6 - General & Particular} we explain why the current traditional Horizontal Position Designs are particular cases of our general PEAL Method.

\subsection{Designing definition} \label{Subsec: Step 6 - Designing Definition}
\begin{figure}[!ht]
	\begin{center}		\includegraphics[width=0.8\textwidth,height=\textheight,keepaspectratio]{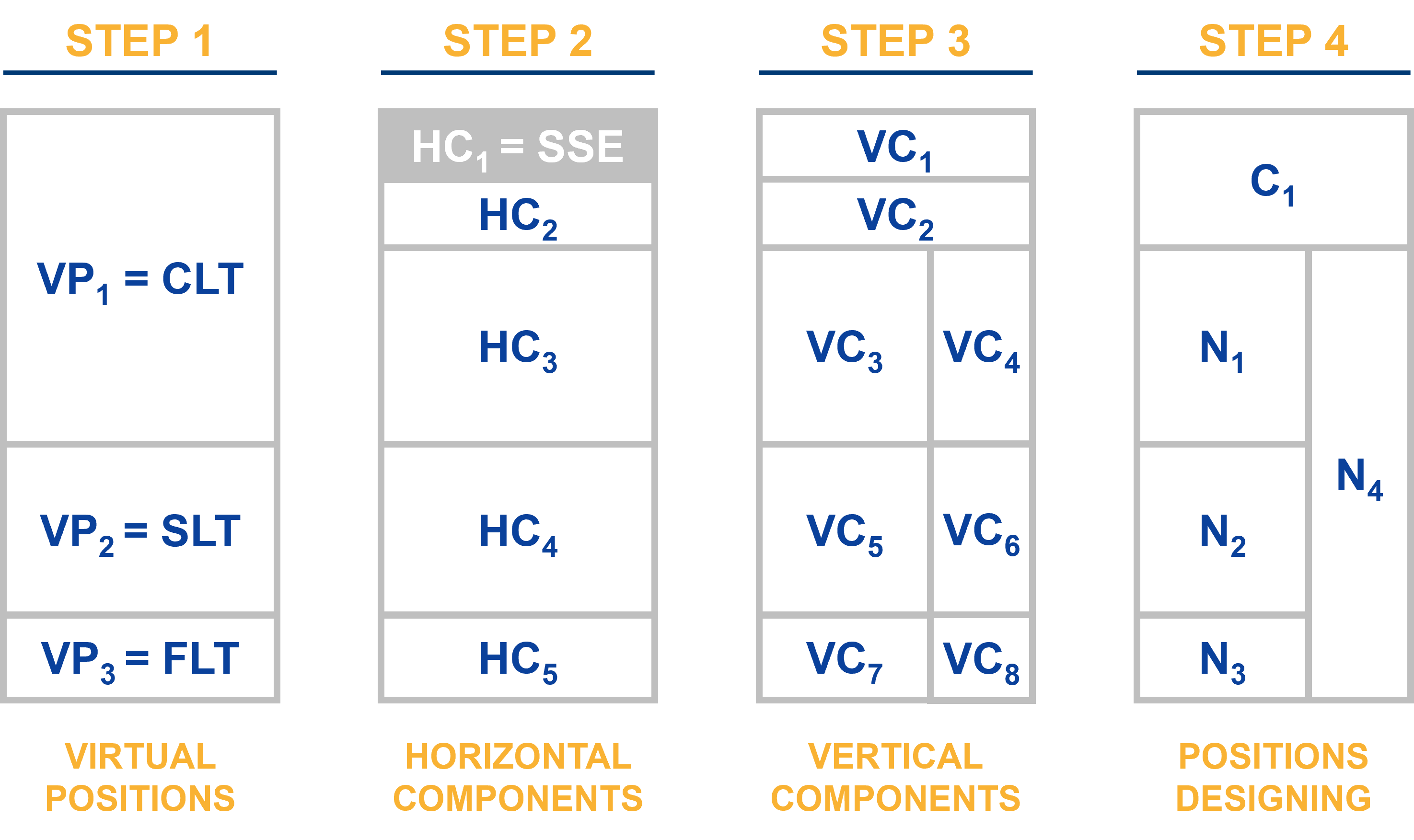}    
	\end{center}
	\caption{Synoptic representation of the 4 steps required to Designing the Positions. In Step 1 we compute the Virtual Positions. In Step 2 we decide the number of Horizontal Components and their $h$ \% weight to their relative Virtual Positions. In Step 3 we decide the number of Vertical Components and their $v$ \% weight to their relative Horizontal Components. Finally, in Step 4 we design the Waterfall Configuration by connecting each $x$ Cost and $y$ Note Position to their Vertical Components. }
	\label{fig: step 6}
\end{figure}

We define the process of organizing the $X$ Cost and the $Y$ Note Positions so as to absorb all the Tranches as \textbf{Designing}, and its synoptic graphical representation as a \textbf{Waterfall Configuration}\footnote{In technical jargon, we refer to a \say{Waterfall} because the cash \say{flows down} from the \say{higher-tier} Position to the \say{lower-tiers} Positions, in a way for which the \say{higher-tier} Positions are paid before the others, following the Waterfall Configuration, and the \say{lowest-tier} Position is paid only if there is any cash flow left.}.
The 6th Step of any Structuring Method is to Design the $P$ Positions into a Waterfall Configuration. The Designing process is made of 4 steps (see Figure \ref{fig: step 6}): 
\begin{enumerate}
	\item \textbf{Virtual Positions}: define a \textit{superstructure} that bridges the inbound and the outbound cash flows;
	\item \textbf{Horizontal Components}: define a \textit{superstructure} that horizontally subdivide, and mathematically links, each Virtual Position in smaller horizontal rectangles;
	\item \textbf{Vertical Components}: define a \textit{superstructure} that vertically subdivides, and mathematically links, each Horizontal Position in smaller rectangles;
	\item \textbf{Positions Design}: define the equations that logically connect each $x$ Cost and the $y$ Note Position to their relative Vertical Components.
\end{enumerate}

To date, there are 3 Waterfall Configurations: horizontal; vertical; and hybrid. In the \textbf{Horizontal Waterfall}, each Position absorbs the Tranching in horizontal slices: the $TAF^{(s)}(t)$ flows down from the top to the bottom Position following the Designing (i.e. the Position $a$ has higher payment priority than the Position $b$ if and only if $a<b$). In the \textbf{Vertical Waterfall}, each Position absorbs the Tranching in vertical slices: the $TAF^{(s)}(t)$ is allocated to the vertical Positions \textit{pro-rata} basis. In the \textbf{Hybrid Waterfall}, each Position absorbs the Tranching in a hybrid combination of horizontal and vertical: the $TAF^{(s)}(t)$ is allocated to the Positions indirectly, through the Virtual Components.

\subsection{Virtual Positions} \label{Subsec: Step 6 - Virtual Positions}
Let a \textbf{Virtual Position} $VP$ be a mathematical \textit{superstructure} that bridges the inbound and the outbound cash flows. Let $NP$ be the number of Virtual Positions (see Step 1 in Figure \ref{fig: step 6}). In this paper, the number of Virtual Positions is fixed at $NP = 3$. Thus, let the system of equations that connect the Virtual Positions outbound cash flows to the inbound cash flows be
\begin{equation} \label{eq: Virtual Position}
	\begin{cases}
		VP_1(t)  = & CLT(t)\\
		VP_2(t)  = & SLT(t)\\
		VP_3(t)  = & FLT(t)
	\end{cases}
\end{equation}
This choice was made to simplify the regulatory capital optimization and computation. Indeed, as explained in the previous Sec. \ref{Subsec: Step 5 - Positions Qualities}, the First Loss Tranche, the Second Loss Tranche, and the Complementary Loss Tranche on the asset-side correspond to the Junior Quality, the Mezzanine Quality, and the Senior Quality on the liability-side. It follows that $VP_1 = Senior \ Quality$, $VP_2 = Mezzanine \ Quality$, and $VP_3 = Junior \ Quality$. 

Being a Horizontal Waterfall type, Virtual Positions are useful because they have a robust mathematical link to the inbound cash flows and introduce a clear priority in the Positions' payments, enabling simpler computation in preparation of the next steps. We will use the Virtual Positions as a pivotal element to compute the net cash flows to the Positions in the different Scenario $s$ in the following Sec. \ref{Sec: Step 8 - Net Dimensioning}.

\subsection{Horizontal Components} \label{Subsec: Step 6 - Horizontal Components}
Let a \textbf{Horizontal Component} $HC$ be a mathematical \textit{superstructure} that horizontally subdivides, and mathematically links, the $np$-th Virtual Position into horizontal rectangles (see Step 2 in Figure \ref{fig: step 6}). Let the set of horizontal slices of the $NP$ Virtual Positions be $HS = \{ H_1, H_2, \dots H_{NP} \}$ and 
\begin{equation} \label{eq: H}
	H = \sum_{np = 1}^{NP} H_{np}
\end{equation}
be the number of total horizontal slices, where $H \ge NP$. Notice that the first Horizontal Component is always the average Super Senior Embedded Position (i.e. $H_1(t) = \overline{SSE}(t)$). Let $j$ be the index of the $j$-th slice (from top to bottom) and let $h_{j}(t)$ be the \textbf{Horizontal Percentage} with respect to its Virtual Position $VP$. As an example, let consider the Step 2 in the previous Figure \ref{fig: step 6} where: $NP = 3$, $HS = \{ 3,1,1 \}$, $H = 5$. We can infer that $h_1(t)$, $h_2(t)$ and $h_3(t)$  \say{slice} $VP_1$, $h_4(t)$ \say{slices} $VP_2$ and $h_5(t)$ \say{slices} $VP_3$. Obviously, we have that $h_1(t)+h_2(t)+h_3(t)=100\%$, $h_4(t)=100\%$, and $h_5(t)=100\%$. Then, the equations that connect the Virtual Positions to the Horizontal Components are
\[
\begin{cases} \label{eq: Horizontal Set}
	HC_1(t)   = & VP_1(t) \cdot h_1(t)\\
	HC_2(t)   = & VP_1(t) \cdot h_2(t)\\
	HC_3(t)   = & VP_1(t) \cdot h_3(t)\\
	HC_4(t)   = & VP_2(t) \cdot h_4(t)\\
	HC_5(t)   = & VP_3(t) \cdot h_5(t)\\
\end{cases}
\]
where $VP_1(t) \cdot h_1(t) = \overline{SSE}(t)$ in any Structuring Method.

\subsection{Vertical Components} \label{Subsec: Step 6 - Vertical Components}
Let a \textbf{Vertical Component} $VC$ be a mathematical \textit{superstructure} that vertically subdivides, and mathematically links, the $j$-th Horizontal Component in smaller rectangles (see Step 3 in Figure \ref{fig: step 6}). Let the set of vertical slices of the $H$ Horizontal Components be $VS = \{ V_1, V_2, \dots V_{H} \}$ and 
\begin{equation} \label{eq: V}
	V = \sum_{j = 1}^{H} V_{j}
\end{equation}
be the number of total rectangles composing the securitization liability-side, where $V \ge H$. Notice that the first Vertical Component cannot be subdivided and it is always equal to the Super Senior Embedded Position (i.e. $H_1(t) = V_1(t) = \overline{SSE}(t)$).

Let $i$ be the index of the $i$-th slice (from top to bottom, from left to right) and let $v_i(t)$ be the \textbf{Vertical Percentages} utilized to subdivide the Horizontal Components into their Vertical Components. As an example, let consider the Step 3 in the previous Figure \ref{fig: step 6} where: $NP = 3$, $HS = \{ 3,1,1 \}$, $H = 5$, $VS = \{ 1,1,2,2,2 \}$, $V = 8$. We can infer that: $v_1(t)$ \say{slices} $HC_1$, $v_2(t)$ \say{slices} $HC_2$, $v_3(t)$ and $v_4(t)$ \say{slice} $HC_3$, $v_5(t)$ and $v_6(t)$ \say{slice} $HC_4$, and  $v_7(t)$ and $v_8(t)$ \say{slice} $HC_5$. It naturally follows that $v_1(t)=100\%$, $v_2(t)=100\%$, $v_3(t)+v_4(t)=100\%$, $v_5(t)+v_6(t)=100\%$, $v_7(t)+v_8(t)=100\%$. Then, the equations that connect the Horizontal Components to the Vertical Components are
\[
\begin{cases} \label{eq: Vertical Set}
	VC_1(t)  = & HC_1(t) \cdot v_1(t)\\
	VC_2(t)  = & HC_2(t) \cdot v_2(t)\\
	VC_3(t)  = & HC_3(t) \cdot v_3(t)\\
	VC_4(t)  = & HC_3(t) \cdot v_4(t)\\
	VC_5(t)  = & HC_4(t) \cdot v_5(t)\\
	VC_6(t)  = & HC_4(t) \cdot v_6(t)\\
	VC_7(t)  = & HC_5(t) \cdot v_7(t)\\
	VC_8(t)  = & HC_5(t) \cdot v_8(t)\\
\end{cases}
\]
Notice that, while the order of the Horizontal Components provides some information on their payment priorities, the order of the Vertical Components do not. Any order is acceptable because it is the arranger that provide the system of equations to link each $j$-th Horizontal Component to the relative $i$-th Vertical Component.

\subsection{Positions' Designing} \label{Subsec: Step 6 - Positions' Designing}
Let the process to mathematically connect each $x$ Cost and $y$ Note Position to their relative $V_i$ Vertical Components be defined as \textbf{Positions' Designing} (see Step 4 in Figure \ref{fig: step 6}). Following on the example described in the previous Sections, we can define the system of equations of Step 4 of the previous Figure \ref{fig: step 6} that connect the Vertical Components outbound cash flows to each $x$ Cost and $y$ Note Position outbound cash flows as
\[
\begin{cases} \label{eq: Design Set}
	C_1(t) = & VC_1(t) + VC_2(t)\\
	N_1(t) = & VC_3(t)\\
	N_2(t) = & VC_5(t)\\
	N_3(t) = & VC_7(t)\\
	N_4(t) = & VC_4(t) + VC_6(t) + VC_8(t)\\
\end{cases}
\]

\subsection{Illustrative example} \label{Subsec: Step 6 - Explanation of the example}
We choose the specific example of a design (Figure \ref{fig: step 6}) to illustrate the utilization of horizontal and vertical components in defining Note Positions. Moreover, the introduction of this type of design could have significant implications in the practice of securitization. 

Reg 2402, article 6(3)(a) requires that certain actors shall retain on an ongoing basis a material net economic interest which, in any event, shall not be less than $5\%$ of the nominal value of each of the Tranches sold or transferred to investors (i.e. a vertical slice of the $Y$ Note Positions). In addition, pursuant article 5(1)(c) of Reg 2402, institutional investors are required to verify that the mandatory retention is maintained on an ongoing basis by the relevant actors. Unfortunately, the operation to verify the continuous maintaining of the mandatory retention is currently extremely impervious, and often times, almost impossible due to operative constraints. Thus, the possibility to create a vertical Note - $N_4$ in step 4 of Figure \ref{fig: step 6} - dimensioned to respect the said $5\%$ mandatory retention, might solve all the current operative issues at once. If $N_4$ be hold by a third-party, then any investor could just request at any point in time the trustee to certify that the $N_4$ vertical Position be in the possession of the relevant actors. This alone might automatically fulfill Reg 2402 due diligence requirements. 

Therefore, the usage of vertical Positions is not just a theoretical or aesthetic choice, but has profound regulatory implications. Indeed, just by creating such Position, it would be possible to minimize the due diligence operative complexity and costs for both the institutional investors and the supervisory authorities. 

\subsection{A particular Design of a general method} \label{Subsec: Step 6 - General & Particular}
As explained in the previous Sec. \ref{Subsec: Step 6 - Designing Definition}, there are 3 type of Waterfall Configurations. Nowadays, the most used Waterfall Configurations pertains of $X = 1$ Cost Position and $Y = 3$ Note Positions.  
Such notes follow the same order of Figure \ref{fig: step 6} where $N_4$ does not exist and $C_1 = HC_1 + HC_2$, $N_1 = HC3$, $N_2 = HC_4$, and $N_3 = HC_5$. Thus, the current typical Horizontal Waterfall Configurations are just particular cases of the PEAL Method general framework, where the $p = 1 \dots P$ Positions are perfectly superimposable to the $i = 1 \dots V$ Vertical Components and to the $j = 1 \dots H$ Horizontal Components. The Designing of Sec. \ref{Sec: Step 6 - Designing}, the Gross Dimensioning of Sec. \ref{Sec: Step 7 - Gross Dimensioning}, and the Net Dimensioning of Sec. \ref{Sec: Step 8 - Net Dimensioning} processes remain the same. 

\section{Step 7: Dimension the Gross Cost \& Gross Notes} \label{Sec: Step 7 - Gross Dimensioning}
In this section we outline the process of gross dimensioning divided into seven steps. In Sec. \ref{Subsec: Step 7 - Gross Dimensioning Definition} we define the process to dimension the Gross Cost $GC$ and Gross Notes $GN$. In Sec. \ref{Subsec: Step 7 - Frequencies} we introduce the concept of payment frequencies, that are crucial in understanding payment waterfalls, and discuss how to avoid incongruities with the specific rules of Reg 2402.  In Sec. \ref{Subsec: Step 7 - FTF} we introduce the frequency transformation function to adjust cash flow vectors into constant intervals. in Sec. \ref{Subsec: Step 7 - GV} we dimension the Gross Vertical Component $GV$; in Sec. \ref{Subsec: Step 7 - GC e GN} we dimension the Gross Cost $GC$ and the Gross Note $GN$; in Sec. \ref{Subsec: Step 7 - GH} we dimension the Gross Horizontal Component $GH$; and finally in Sec. \ref{Subsec: Step 7 - Horizontal Rule g-check} we provide a simple test to check if we are correctly applying the \say{horizontal rule} of Reg 2402. 

\subsection{Gross Dimensioning definition} \label{Subsec: Step 7 - Gross Dimensioning Definition}
We define the process to indirectly dimension the Gross Cost $GC(t)$ and the Gross Note $GN(t)$ following the system of equations described in the previous Designing phase, putting their monthly cash flows at their relevant payment frequency, as \textbf{Gross Dimensioning}. The 7th Step in any Structuring Method is to compute $GC(t)$ and $GN(t)$. The Gross Dimensioning process is divided in 5 steps:
\begin{enumerate}
	\item \textbf{Frequencies}: select the Vertical Components' payment frequencies;
	\item \textbf{GV}: compute the Vertical Component Gross Dimensioning;
	\item \textbf{GC} \& \textbf{GN}: compute Costs' and Notes' Gross Dimensioning;
	\item \textbf{GH}: compute the Horizontal Components' Gross Dimensioning;
	\item \textbf{g-check}: compute the $g$ ratio to check for coherence with Reg 2402.
\end{enumerate}

\subsection{Select the frequencies} \label{Subsec: Step 7 - Frequencies}
In any Structuring Method, the frequencies are the \say{keystones} to understand the functioning of the payment waterfalls. In a pure mathematical modelling perspective, any Vertical Component could have its own payment frequency $f(VC)$, without considering its relation with the others and without any regulatory restriction. In practice, there are some constraints that shall be respected in selecting the payment frequencies. Reg 2402 article 2(1)(b), requires that the distribution of losses on the ongoing life of the transaction be dependent on the subordination of the Tranches. In order to respect such prescription, the payment frequency must follow 3 rules:
\begin{itemize}
	\item \textbf{vertical}: the \say{higher} Horizontal Components must have an higher payment frequency than the \say{lower} Horizontal Components (i.e. $f(HC_1) \ge f(HC_2) \ge \dots \ge f(HC_{H})$); 
	\item \textbf{multiple}: the \say{lower} Horizontal Components must have a frequency that is a multiple of the \say{highest} Horizontal Component (i.e. $ f(HC_i) \equiv 0 \pmod{f(HC_{i+1})}$;
	\item \textbf{horizontal}: the Vertical Components that belong to the same Horizontal Component must have the same frequency (i.e. $f(VC_a) = f(VC_b)$ if $\{ VC_a, VC_b \} \in HC$.    
\end{itemize}

If we relax even just one of the above rules, some of the Features of the next Sec. \ref{Sec: Step 9 - Features} will show signs of incongruities. For example if we relax the \say{vertical rule} and we pay a \say{lower-tier} Horizontal Component (e.g. the $j$-th $HC_j \in FLT$) with a higher frequency than the \say{higher-tier} Horizontal Component (e.g. the $j$-th $HC_j \in CLT$), then in case of stressed scenarios the \say{higher-tier} might be less protected than the \say{lower-tier}, and thus bear more losses than what it would have done otherwise, contradicting the Reg 2402 prescriptions. If we relax the \say{horizontal rule} and pay two Vertical Components that are virtually \say{pari-passu} with two different frequencies, we are making the part that is paid with a higher frequency less risky than the other on an ongoing basis, thus contradicting the Reg 2402. Finally, even if we respect both the \say{vertical} and \say{horizontal} rules, if we relax the \say{multiple} rule, then in certain periods we might be providing protection to a \say{lower-tier} Horizontal Component whose cash flows should have been used in the next periods to protect the \say{higher-tiers} from negative events, contradicting Reg 2402. 

As we will further explain, we will use the Coefficient of Variation Adjusted $CVA$, described in the next Sec. \ref{Sec: Step 9 - Features}, to detect breaches of Reg 2402 requirements: indeed, if any Horizontal Component line were to intersect during the ongoing life of the transaction it would mean that at that time a \say{lower-tier} Horizontal Component was less risky than a \say{higher-tier} Horizontal Component, contradicting Reg 2402 requirements. 

\subsection{Frequency transformation function } \label{Subsec: Step 7 - FTF}
Since typically payments to the Positions have a lower frequency than the payments on the asset side, we need a function to transform the monthly inbound cash flows to the Position respective outbound cash flows frequencies. Let $cf(t)$ be the monthly cash flow vector to be transformed to a frequency $\omega_{cf}$ paid each $\tau_{cf} = 12 / \omega_{cf}$ months. Then the transformation
\begin{equation} \label{eq: fcf}
	\mathbb{F}(cf(t),\omega_{cf}) =
	\begin{cases}
		\displaystyle{\sum_{t'=t-\tau_{cf}+1}^{t}} cf(t') & \text{if $t = i \cdot \tau_{cf}$ with $i\in \mathbb{N}$} \\
		0 & \text{otherwise}
	\end{cases}
\end{equation} 

\subsection{Compute GV} \label{Subsec: Step 7 - GV}
Let $GV_i(t)$ be the Gross Dimensioning of the $i$-th Vertical Component as
\begin{equation} \label{eq: GV}
	GV_i(t) = \mathbb{F}(VC_i(t),\omega_{VC_i})
\end{equation}
where $i = 1 \dots V$, $\mathbb{F}$ is defined as Equation \eqref{eq: fcf} and $\omega_{VC_i}$ is the frequency of the outbound cash flows of the $i$-th Vertical Component $VC_i(t)$.

\subsection{Compute GC \& GN} \label{Subsec: Step 7 - GC e GN}
Let $GC_x(t)$ be the Gross Dimensioning of the $x$-th Cost Position as the sum of its Gross Dimensioned Vertical Components. Let $GN_y(t)$ be the Gross Dimensioning of the $y$-th Note Position as the sum of its Gross Dimensioned Vertical Components. Therefore, if we consider the $x$ Cost and $y$ Note Positions of the example of Sec. \ref{Subsec: Step 6 - Positions' Designing}, the result is
\[
\begin{cases} \label{eq: GC & GN}
	GC_1(t) = & GV_1(t) + GV_2(t)\\
	GN_1(t) = & GV_3(t)\\
	GN_2(t) = & GV_5(t)\\
	GN_3(t) = & GV_7(t)\\
	GN_4(t) = & GV_4(t) + GV_6(t) + GV_8(t)\\
\end{cases}
\]
where $GV_i(t)$ is function \eqref{eq: GV}. 

\subsection{Compute GH} \label{Subsec: Step 7 - GH}
Let $GH_j(t)$ be the Gross Dimensioning of the $j$-th Horizontal Component as the sum of its Gross Dimensioned Vertical Components. Thus, if we consider the $H$ Horizontal Components of the example of Sec. \ref{Subsec: Step 6 - Vertical Components}, the result is 
\[
\begin{cases} \label{eq: GH}
	GH_1(t) = & GV_1(t)\\
	GH_2(t) = & GV_2(t)\\
	GH_3(t) = & GV_3(t) + GV_4(t)\\
	GH_4(t) = & GV_5(t) + GV_6(t)\\
	GH_5(t) = & GV_7(t) + GV_8(t)\\
\end{cases}
\]
where $GV_i(t)$ is function \eqref{eq: GV}. 

\subsection{Horizontal rule g-check} \label{Subsec: Step 7 - Horizontal Rule g-check}
Let $g_i(t)$ be the ratio of the $i$-th Gross Vertical to the $j$-th Gross Horizontal Component
\begin{equation} \label{eq: gi}
	g_i(t) = \frac{GV_i(t)}{GH_j(t)}  
\end{equation}
with $GV_i(t) \in GH_j(t)$ as described in each system of equations that will be used in the relative Structuring Method, as shown in the example of the previous Sec. \ref{Subsec: Step 7 - GH}. The $g$-check is a simple mechanism through which we immediately know if the payment frequencies we chose are respecting the 3 frequency rules explained in the previous Sec. \ref{Subsec: Step 7 - Frequencies}. This test states that if the $i$-th gross ratio $g_i(t)$ of a certain Gross Vertical Component $GV_i(t)$ to its $j$-th Gross Horizontal Component $GH_j(t)$ is equivalent to the $v_i(t)$ percentage used to transform the monthly $j$-th Horizontal Component $HC_j(t)$ into its monthly $i$-th Vertical Component $VC_i(t)$ (i.e. $g_i(t) = v_i(t) \ \forall \ t$) then the selected payment frequencies are respecting the \say{horizontal rule} of Reg 2402 prescriptions. Otherwise, if $g_i(t) \neq v_i(t)$ for any $t$ then it means that one or more Gross Vertical Components belonging to the same Horizontal Component do not have the same frequency. Therefore, although they shall be \say{pari-passu}, they are partially not, thus breaching Reg 2402.   

\section{Step 8: Dimension the Net Costs \& Net Notes} \label{Sec: Step 8 - Net Dimensioning}
This Section is divided in six parts: in Sec. \ref{Subsec: Step 8 - Net Dimensioning Definition} we define the process to dimension the Net Cost $NC$ and Net Notes $NN$; in Sec. \ref{Subsec: Step 8 - Gross Dimensioning Matrix} we introduce the Gross Dimensioning Matrix $GDM$; in Sec. \ref{Subsec: Step 8 - Net Dimensioning Matrix} we introduce the waterfall payment function $fwp$ used to compute the Net Dimensioning Matrix $NDM$; in Sec. \ref{Subsec: Step 8 - NV} we dimension the Net Vertical Component $NV$; in Sec. \ref{Subsec: Step 8 - NC & NN} we dimension the Net Cost $NC$ and the Net Notes $NN$; and in Sec. \ref{Subsec: Step 8 - LC & LN} we compute the Cost Losses $LC$ and the Note Losses $LN$. 

\subsection{Net Dimensioning Definition} \label{Subsec: Step 8 - Net Dimensioning Definition}
In a Scenario $s$, we define the indirect process to dimension the Net Cost $NC^{(s)}(t)$ and the Net Notes $NN^{(s)}(t)$ by computing the Net Dimensioning Matrix $NDM^{(s)}(t)$, allocating the total available funds $TAF^{(s)}(t)$ on the Gross Dimensioning Matrix $GDM(t)$ using the Waterfall Payment Function $fwp$, as \textbf{Net Dimensioning}. The 8th Step of any Structuring Method is to compute the Net Dimensioning of the Cost $NC^{(s)}(t)$ and the Note $NN^{(s)}(t)$ Positions. The Net Dimensioning process is divided in 5 steps:
\begin{enumerate}
	\item \textbf{GDM}: define the Gross Dimensioning Matrix;
	\item \textbf{NDM}: compute the Net Dimensioning Matrix;
	\item \textbf{NV}: compute the Vertical Components Net Dimensioning;
	\item \textbf{NC} \& \textbf{NN}: compute the Cost and Note Net Dimensioning;
	\item \textbf{LC} \& \textbf{LN}: compute the Cost and Note Losses.
\end{enumerate}

\subsection{Gross Dimensioning Matrix} \label{Subsec: Step 8 - Gross Dimensioning Matrix}
Let the Gross Dimensioning Matrix $GDM(t)$ be the matrix whose columns follow the $GH_j(t)$ as $GDM(t) = [ \ GH_1(t), \ GH_2(t), \ \dots, \ GH_{H}(t) \ ]$, where $j = 1 \dots H$. Notice that, the GDM is a pure Horizontal Configuration: as it will be shown in the next Sec. \ref{Subsec: Step 8 - Net Dimensioning Matrix}, the Waterfall Payment Function $fwp$ allocates, on an ongoing basis, the total available funds $TAF^{(s)}(t)$ from left to right: thus, the first column on the left is the first Position to have a claim on the total available funds, the second column has the second claim and so forth until the last column $H$ whom has the last claim on any residual liquidity. Therefore, in order to perfectly match the Horizontal Components Designing described in the selected Waterfall Configuration, the Super Senior Embedded Position $SSE$ must always be the first column because it has anyway the first claim on any available funds. Then, the following columns, should be the other Gross Dimensioned Horizontal Components in an ascending order: indeed, we know that $GH_1(t)$ has always an higher claim than $GH_2(t)$ that has an higher claim on $GH_3(t)$ and so forth to $GH_{H}(t)$.

\subsection{Net Dimensioning Matrix} \label{Subsec: Step 8 - Net Dimensioning Matrix}
Let the Waterfall Payment Function be the function that, in a Scenario $s$, allocates the inbound total available funds $TAF^{(s)}(t)$ on the Gross Dimensioning Matrix $GDM(t)$ to respect their loss allocation Designing and their Gross Dimensioning, thus computing the Positions' Net Horizontal Dimensioning $NH^{(s)}(t)$ synthesized into the Net Dimensioning Matrix $NDM^{(s)}(t)$. The Waterfall Payment Function is a system of recursive equations that loops from the top left to the bottom right of the GDM. The loop system is not obvious, and thus we decided to sub-divide it into 3 sequential steps:
\begin{enumerate}
	\item \textbf{Step 1}: compute the total gross position $TGP(t)$;
	\item \textbf{Step 2}: compute the total net position $TNP^{(s)}(t)$;
	\item \textbf{Step 3}: compute the Net Dimensioning Matrix $NDM^{(s)}(t)$.
\end{enumerate}
In synthesis, in any Scenario $s$, we first compute the total net position $TNP^{(s)}(t)$, allocating the total available funds $TAF^{(s)}(t)$ (plus eventual residual cash flows from previous periods) on the total gross position $TGP(t)$ (plus eventual debts from previous periods). Then, we compute the Net Dimensioning Matrix $NDM^{(s)}(t)$ by allocating the total net position $TNP^{(s)}(t)$ payments to each Horizontal Component according to its payment priorities.

\subsubsection{Step 1: compute the total gross position} \label{Subsubsec: step 8 - tgp}
Let the sum of the $j = 1 \dots H$ Gross Dimensioned Horizontal Component $GH_j(t)$ be the total gross position $TGP(t)$ as
\begin{equation} \label{eq: tgp}
	TGP(t) = \sum_{j=1}^H GH_j(t) = \sum_{j=1}^H GDM_j(t)
\end{equation}

\subsubsection{Step 2: compute the total net position} \label{Subsubsec: Step 8 - tnp}
In any Scenario $s$, in each period $t$, let the allocation of the inbound total available funds $TAF^{(s)}(t)$ to the total gross position $TGP(t)$ be
\begin{equation} \label{eq: tnp system} 
	\begin{cases}
		PAF^{(s)}(t) = & TAF^{(s)}(t)+ADV^{(s)}(t-1)\\
		TAD^{(s)}(t) = & TGP(t)+DBT^{(s)}(t-1) \cdot h(\cdot)\\
		ADV^{(s)}(t) = & \max(PAF^{(s)}(t)-TAD^{(s)}(t),0)\\
		DBT^{(s)}(t) = & \max(TAD^{(s)}(t)-PAF^{(s)}(t),0) \cdot h(\cdot) + DBT^{(s)}(t-1) \cdot (1 - h(\cdot))
	\end{cases}
\end{equation}
that is a set of recursive Equations where $h(\cdot)$ is $h(TGP(t))$ and represents an indicator of \emph{when} payments occur: in fact, $h(\cdot)$ is $1$ if and only if any outbound payment to any Position is due at a given period $t$ as follow
\begin{equation} \label{eq: hx}
	h(x)=
	\begin{cases}
		0 & \text{if $x \leq 0 $} \\
		1 & \text{if $x > 0 $}
	\end{cases}
\end{equation} 
Notice that in a generic Scenario $s$ the periodical total available funds $PAF^{(s)}(t)$ can be in general larger than the total available funds $TAF^{(s)}(t)$, since there can be cash advances from the previous period. A simple example is when the Positions are paid starting from the second period: in this case, introducing a dummy variable $ADV^{(s)}(t)$ accounting for the total cash advances, $PAF^{(s)}(2)=TAF^{(s)}(2)+ADV^{(s)}(1)$. On the same footing, we have to introduce a dummy variable $DBT^{(s)}(t)$ accounting for the fact that, if in a payment period $t = 1$ there is not enough cash to pay for the total gross position $TGP(1)$, then in the following payment period $t = 2$ we consider what is due to the positions is $TGP(2)$ \emph{plus} what was not paid at $t = 1$. Thus, we indicate with $TAD^{(s)}(t)$ the total amount due to all the Positions. We have explicitly considered that if at period $t$ there is a payment, then the difference $PAF^{(s)}(t)-TAD^{(s)}(t)$ between available and due can either increase $ADV^{(s)}(t)$ or $DBT^{(s)}(t)$ according to the sign. Notice that it is also necessary to specify the initial value of the dummy variables; the natural choice is $ADV^{(s)}(0)=0$, $DBT^{(s)}(0)=0$. \textbf{NB}: $TAD^{(s)}(t)$ is different from zero \emph{only} at the payment periods. Finally, let the minimum between the periodical total available funds $PAF^{(s)}(t)$ and the total amount due to all the Positions $TAD^{(s)}(t)$ be
\begin{equation} \label{eq: tnp}
	TNP^{(s)}(t) = \min(PAF^{(s)}(t),TAD^{(s)}(t))
\end{equation}
where the total net position $TNP^{(s)}(t)$ takes into account the maximum amount of periodical inbound cash flows that can be allocated to all the $H$ Positions at any given period $t$. Notice that $\sum_{t=0}^{TP} TNP^{(s)}(t) = \sum_{t=0}^{TP} TAF^{(s)}(t)$.

\subsubsection{Step 3: compute the net dimensioning matrix} \label{Subsubsec: Step 8 - NDM}
In any Scenario $s$, in each period $t$, let the Net Dimensioning Matrix $NDM^{(s)}(t)$ be the matrix whose columns are composed of the $j = 1 \dots H$ Net Dimensioned Horizontal Components $NH_j(t)$, computed using the following system of recursive equations 
\begin{equation} \label{eq: NDM}
	\begin{cases}
		NDM_j^{(s)}(t)=\min\left(GDM_j(t)+DBT_j^{(s)}(t-1),RNP_{j-1}^{(s)}(t)\right) &\\
		DBT_j^{(s)}(t) = \min\left(0, GDM_j(t) - NDM_j^{(s)}(t) \right) \cdot h(\cdot) + DBT_j^{(s)}(t-1) \cdot (1- h(\cdot)) & \\
		RNP_j^{(s)}(t)= RNP_{j-1}^{(s)}(t)-NDM_j^{(s)}(t)  &\\
	\end{cases}
\end{equation}
where, for compactness, we indicate $h(TGP_j(t))$ as $h(\cdot)$. The expression for $NDM_j^{(s)}(t)$ in Equation \eqref{eq: NDM} ensures that any $j$-th Horizontal Component cannot receive more than their respective $TGP_j(t)$ plus any eventual debt $DBT_j^{(s)}(t-1)$ of the previous period, and at most the cash amount $RNP_{j-1}^{(s)}(t)$ residual after paying the Horizontal Component $j-1$. The expression for $DBT_j^{(s)}(t)$ accounts for the eventual debt from the previous period and for the difference between the due amount $GDM_j(t)$ and the net amount that has really been paid net of any loss $NDM_j^{(s)}(t)$. 

We recall that $h(GDM_j(t))$ indicates whether any payment to the $j$-th Horizontal Component is occurring at period $t$. $RNP_j^{(s)}(t)$ accounts for the fact that the residual cash for $j$-th Horizontal Component equals the difference between the residual cash for the Horizontal Component $j-1$ and what has been paid to that Horizontal Component. Also in this case it is necessary to specify the initial values of the dummy variables $RNP_{j=0}^{(s)}(t)=TNP^{(s)}(t)$ and $DBT_j^{(s)}(0)=0$. Notice that $\sum_{j=1}^{H} NDM_j^{(s)}(t) = TNP^{(s)}(t)$.

\subsection{Vertical Components Net Dimensioning} \label{Subsec: Step 8 - NV}
Let the product of $j$-th Net Dimensioned Horizontal Components $NDM_j(t)$ per their respective gross ratio $g_i$ be the Vertical Components Net Dimensioning $NV_i^{(s)}(t)$
\begin{equation} \label{eq: nv}
	NV_i^{(s)}(t) = NDM_j^{(s)}(t) \cdot g_i
\end{equation}
where $g_i$ is Equation \eqref{eq: gi}. So, given the example in Sec. \ref{Subsec: Step 6 - Vertical Components} the result is
\[
\begin{cases} \label{eq: nv example}
	NV^{(s)}_1(t)  = & NDM^{(s)}_1(t) \cdot g_1(t)\\
	NV^{(s)}_2(t)  = & NDM^{(s)}_2(t) \cdot g_2(t)\\
	NV^{(s)}_3(t)  = & NDM^{(s)}_3(t) \cdot g_3(t)\\
	NV^{(s)}_4(t)  = & NDM^{(s)}_3(t) \cdot g_4(t)\\
	NV^{(s)}_5(t)  = & NDM^{(s)}_4(t) \cdot g_5(t)\\
	NV^{(s)}_6(t)  = & NDM^{(s)}_4(t) \cdot g_6(t)\\
	NV^{(s)}_7(t)  = & NDM^{(s)}_5(t) \cdot g_7(t)\\
	NV^{(s)}_8(t)  = & NDM^{(s)}_5(t) \cdot g_8(t)\\
\end{cases}
\]

\subsection{Cost \& Note Net Dimensioning} \label{Subsec: Step 8 - NC & NN}
Let $NC^{(s)}_x(t)$ be the Net Dimensioning of the $x$-th Cost Position as the sum of its Net Dimensioned Vertical Components. Let $NN^{(s)}_y(t)$ be the Net Dimensioning of the $y$-th Note Position as the sum of its Net Dimensioned Vertical Components. Therefore, if we consider the $x$ Cost and $y$ Note Positions of the example of Sec. \ref{Subsec: Step 6 - Positions' Designing}, the result is
\[
\begin{cases} \label{eq: NC & NN}
	NC^{(s)}_1(t) = & NV^{(s)}_1(t) + NV^{(s)}_2(t)\\
	NN^{(s)}_1(t) = & NV^{(s)}_3(t)\\
	NN^{(s)}_2(t) = & NV^{(s)}_5(t)\\
	NN^{(s)}_3(t) = & NV^{(s)}_7(t)\\
	NN^{(s)}_4(t) = & NV^{(s)}_4(t) + NV^{(s)}_6(t) + NV^{(s)}_8(t)\\
\end{cases}
\]
where $NV^{(s)}_i(t)$ is function \eqref{eq: nv}. 

\subsection{Cost \& Note Losses} \label{Subsec: Step 8 - LC & LN}
Let $LC_x^{(s)}(t)$ be the Loss of the $x$-th Cost Position as the difference between the Gross Dimensioned Cost Position $GC_x^{(s)}(t)$ and the Net Dimensioned Cost Position $NC_x^{(s)}(t)$ as
\begin{equation} \label{eq: lc}
	LC_x^{(s)}(t) = GC_x^{(s)}(t) - NC_x^{(s)}(t)
\end{equation}
Let $LN_y^{(s)}(t)$ be the Loss of the $y$-th Note Position as the difference between the Gross Dimensioned Note Position $GN_y^{(s)}(t)$ and the Net Dimensioned Note Position $NN_y^{(s)}(t)$ as
\begin{equation} \label{eq: ln}
	LN_y^{(s)}(t) = GN_y^{(s)}(t) - NN_y^{(s)}(t)
\end{equation}
Notice that both the Cost and Note Losses amounts and probability density functions are information often not shared with the stakeholders, although many institutional investors would appreciate to have access to these forecasting to hedge their Positions from excessive risk. Therefore, as with the other securitization Features computed in the next Sec. \ref{Sec: Step 9 - Features}, these key performance indicators shall always be computed and made available to the interested third parties for the ongoing life of the securitization.

\section{Step 9: Compute the relevant Features} \label{Sec: Step 9 - Features}
As introduced in Sec. \ref{Subsec: Step 8 - LC & LN}, there are a number of securitization key performance indicators that the servicers should always provide to the investors on an ongoing basis (\textbf{Features}) to enable better and more transparent risk assessments. The 9th Step of any Structuring Method is to compute, in every Scenario $s$, all the relevant Features. The following is a non-exhaustive list of some of the main Features that we strongly believe the servicers should always provide to the stakeholders on an ongoing basis:
\begin{itemize}
	\item $EP$: the $N$ Exposures Performance (Sec. \ref{Subsec: Step 9 - Performance});
	\item $TH$: the Positions' Thickness (Sec. \ref{Subsec: Step 9 - Position Thickness});
	\item $RC$: the Regulatory Capital (Sec. \ref{Subsec: Step 9 - Regulatory Capital});
	\item $CVA$: the Coefficient of Variation Adjusted (Sec. \ref{Subsec: Step 9 - CVA});
	\item $FV$: the Fair Value (Sec. \ref{Subsec: Step 9 - Fair Value});
	\item $gr$: the gross internal rate of return (Sec. \ref{Subsec: Step 9 - IRR}).
\end{itemize}
Inferring the Features statistics by Monte Carlo simulations allows for a better understanding of the real securitization ongoing intrinsic riskiness. It becomes easier to make all those advanced analyses that note-holders may need to demonstrate to their supervisory authorities that they have done proper risk due diligence during the purchasing and, afterwards, in the ongoing management process. Indeed, the PEAL Method increases the transparency that, in turn, might attract new investors and revive the securitization industry that, since 2008, has shown little sign of resilience. In fact, the European Corporate Loan niche alone lost more than 50\% of new issuance between 2008 and 2017 \citep{kraemer2018} and has never recovered since.

\subsection{Exposure Performance (EP)} \label{Subsec: Step 9 - Performance}
Let the Exposure Performance $EP_{k,n}^{(s)}$ be the function that describes the \say{performance} of each Exposure $(k,n)$ in any Scenario $s$ assigning a Mutually Exclusive and Commonly Exhaustive \say{state}. The 4 possible states are: $\{$Full-Performing, Performing, Non-Performing, Super-Performing$\}$. The $(K,N)$ Exposure Performance is the $1$-st Feature that shall be provided to the securitization stakeholders on an ongoing basis. 

When applying the IAS-IFRS accounting principles, Exposures mainly exist in 3 \say{states}: (i) performing; (ii) unlikely-to-pay (\textbf{UTP}); and (iii) non-performing (\textbf{NPE}). This typification does not provide any information on the real performance of the underlying Exposures in the different Scenarios $S$ versus the Base Scenario. For the purpose of standardization, we propose that the definition of \emph{performing} and \emph{non-performing} be shifted from the current IAS-IFRS accounting principles to the Exposures Performance $EP$, using the following Mutually Exclusive and Commonly Exhaustive function
\begin{equation} \label{eq: EP}
	EP_{k,n}^{(s)} = \sum_{t=0}^{TP} \frac{[R_{k,n}^{(s)}(t) - R_{k,n}^{(b)}(t)] + E_{k,n}^{(s)}(t) - [CR_{k,n}^{(s)}(t) + ER_{k,n}^{(s)}(t)] }{(1+\eta_t)^t}
\end{equation}
where $R_{k,n}^{(s)}(t)$ is Equation \eqref{eq:Rskn ext}; $R_{k,n}^{(b)}(t)$ is Equation \eqref{eq:Rbkn}; $E_{k,n}^{(s)}(t)$ is Equation \eqref{eq: eskn}; $CR_{k,n}^{(s)}(t)$ and $ER_{k,n}^{(s)}(t)$ are components of Equation \eqref{eq: SSE}; and finally $\eta$ is the annual expected inflation rate (e.g. $\eta = 2\% $) while $\eta_t$ is the monthly inflation rate (e.g. $\eta / 12$). Although some Events per se might be inherently negative (e.g. Exposure defaults) or positive (e.g. Exposure return to life), when more than one Event jointly affects the Exposure at different times $\hat{t}_{k,n}^{(\lambda)}$ in a specific Scenario $s$, it becomes extremely difficult \textit{ex-ante} to define if the overall \textit{ex-post} outcome. Considering that every Event materializes at a certain time $\hat{t}_{k,n}^{(\lambda)}$, where $0 \leq \hat{t}_{k,n}^{(\lambda)} \leq T_{k,n}$ (i.e. $\hat{t}_{k,n}^{(\lambda_1)}$, $\hat{t}_{k,n}^{(\lambda_2)}$, $\dots$), then
\begin{equation} \label{eq:ts}
	\hat{t}_{k,n}^{(s)} = min (\hat{t}_{k,n}^{(\lambda_1)}, \hat{t}_{k,n}^{(\lambda_2)}, \dots)
\end{equation}
is the first time when any Event affects the Exposure $(k,n)$ in Scenario $s$. In a Scenario $s$ let an Exposure $(k,n)$ be \textbf{Performing} in the range $r = \hat{\tau}_k \leq t \leq \hat{t}_{k,n}^{(s)}$: indeed Equation \eqref{eq: EP} is always 0 because when an Exposure is Performing it means that its paying exactly as expected in the Base Scenario and thus $E(t) = CR(t) = ER(t) = 0 \ \forall \ t \in r$ and $R^{(b)}(t) - R^{(s)}(t) = 0 \ \forall \ t \in r$. In a Scenario $s$ let an Exposure $(k,n)$ be \textbf{Full-Performing} if $\hat{t}_{k,n}^{(s)} = TP$. In a Scenario $s$ let an Exposure $(k,n)$ be \textbf{Non-Performing} when $t \geq \hat{t}_{k,n}^{(s)}$ if $OP < 0$. Finally, in a Scenario $s$ let an Exposure $(k,n)$ be \textbf{Super-Performing} when $t \geq \hat{t}_{k,n}^{(s)}$ if $OP > 0$. Notice that, even Exposures that are natively $UTP$ or $NPL$ under the IAS-IFRS, in the PEAL Method can still be considered Fully-Performing, Performing or Super-Performing in the different Scenarios $s$ if they follow the rules of Equation \eqref{eq: EP}.  

The novel approach of this Section typifies the Exposures to provide concrete and timely information to stakeholders. Indeed, it is not irrelevant to make advanced risk analyses to properly communicate \textit{overtime} the  mean, the median, and the estimated likelihood of the number of Exposures that, in the various relevant Scenario $S$, are expected to be Full-Performing, Performing, Non-Performing, or Super-Performing.

\subsection{Position Thickness (TH)} \label{Subsec: Step 9 - Position Thickness}
In this Section we will explain the regulatory definition of the $p$-th Position Thickness $TH_p(t)$, and then provide a simplified definition within the PEAL Method that is substantially equal, although formally different, as the ones of the European Regulations. The Position Thickness is the $2$-nd Feature that shall be provided to the securitization stakeholders on an ongoing basis.    

\subsubsection{The Thickness regulatory definition} \label{Subsubsec: Step 9 - Thickness Regulatory Definition}
Reg 2401, Article 263(5), defines the $p$-th Position \textit{thickness} percentage as
\begin{equation} \label{eq:th percentage}
	THP_p(t) = DP_p(t) - AP_p(t)
\end{equation}
where $p = 1 \dots P$ and where, pursuant Article 256 of Reg 2401, $DP_p(t)$ is the detachment point of Position $p$; and $AP_p(t)$ is the attachment point of Position $p$. Reg 2401 implicitly assumes that the $p = 1 \dots P$ Positions be perfectly superimposable at the $j = 1 \dots H$ Horizontal Components, and thus that any Design the arranger could come up with would have no Vertical Positions. Such an hidden assumption not only is extremely strong, but we have already debunked it in the example of the previous Figure \ref{fig: step 6}. Therefore, it becomes mandatory to provide an alternative solution to the computation of the Positions' Thickness as the one proposed in the next Sec. \ref{Subsubsec: Step 9 - Thickness PEAL Method}.  

Reg 2401, Article 256(1), defines the \textit{attachment point} of the $p$-th Position $AP_p(t)$ as \say{the threshold at which losses within the pool of underlying exposures would start to be allocated to the relevant securitization position. The attachment point shall be expressed as a \textit{decimal value between zero and one} and shall be equal to the greater of zero and the ratio of the outstanding balance of the pool of underlying exposures in the securitization minus the outstanding balance of all tranches that rank senior or pari-passu to the tranche containing the relevant securitization position including the exposure itself to the outstanding balance of all the underlying exposures in the securitization}. Then
\begin{equation} \label{eq:apt}
	AP_p(t) = \frac{OBP(t) - \sum_{i=1}^p OB_i(t)}{OBP(t)} = \frac{\sum_{i=p+1}^P OB_i(t)}{OBP(t)}
\end{equation}
is the attachment point, where
\begin{equation} \label{eq:obp}
	OBP(t) = \sum_{p=1}^{P} OB_p(t)
\end{equation}
is the total outstanding balance, and where 
\begin{equation} \label{eq:obpt}
	OB_p(t) = \sum_{t=0}^{TP} GDM_p(t) - \sum_{\tau=0}^t GDM_p(\tau)
	= \sum_{\tau= t+1}^{TP} GDM_p(\tau)
\end{equation}
is the outstanding balance of the each $p$-th Position. Remember that $AP(t)_p \in \{ 0,1 \}$. Notice that $GDM_p(t)$ represents the $p$-th column of the Gross Dimensioning Matrix as explained in the previous Sec. \ref{Subsec: Step 8 - Gross Dimensioning Matrix} and that this formula works if and only if the $p = 1 \dots P$ Positions are perfectly superimposable to their Horizontal Components (i.e. the Waterfall Configuration is composed of only horizontal slices). In the moment that the Positions were vertical or had other shapes, the Regulatory approach would fail and the PEAL Method general equation would be required (see next Sec. \ref{Subsubsec: Step 9 - Thickness PEAL Method}). 

On the other hand, Reg 2401, Article 256(2), defines the \textit{detachment point} of the $p$-th Position $DP_p(t)$ as the \say{threshold at which losses within the pool of underlying exposures would result in a complete loss of principal for the tranche containing the relevant securitization position. The detachment point shall be expressed as a \textit{decimal value between zero and one} and shall be equal to the greater of zero and the ratio of the outstanding balance of the pool of underlying exposures in the securitization minus the outstanding balance of all tranches that rank senior to the tranche containing the relevant securitization position to the outstanding balance of all the underlying exposures in the securitization}. The most relevant aspect to stress about the above Article is that it defines the detachment point as \say{the loss $[...]$ would result in a complete loss of principal $[...]$ for the relevant securitization position}. Indeed, as it happens in all financial products, the losses are firstly absorbed by the interest and only then by the capital. It follows that, in order to have a \say{complete loss of the principal}, the relevant $p$-th Position must have absorbed a loss whose amount is the sum of its Gross Dimensioned interest and capital, represented by $GDM_p(t)$. It might seem strange that we stress this out, but even if this fact is lapalissian, there have been situations in which the arrangers have calculated the $p$-th Position detachment point using only the outstanding capital. This is particularly relevant when calculating the Regulatory Capital of those Positions absorbing the First Loss Tranche $FLT(t)$, for which purposely neglecting to consider the interests considerably diminishes the overall computation (see next Sec. \ref{Subsec: Step 9 - Regulatory Capital}). We want to point out that this blatantly incorrect practice provides a false representation of the Positions' intrinsic riskiness to the detriment of unaware investors. Then
\begin{equation} \label{eq:dpt}
	DP_p(t) =
	\begin{cases}
		1 & \text{if $p = 1$} \\
		\frac{OBP(t) - \sum_{j=1}^{p-1} OB_j(t)}{OBP(t)} = AP_{p-1}(t) & \text{if $p > 1$}
	\end{cases}
\end{equation}
is the detachment point of the $p$-th Position. Remember that $DP_p(t) \in \{ 0,1 \}$. Notice that each $DP_p(t)$ is obviously the $AP_p(t)$ of Position $(p-1)$ and the Position $p = 1$ can fully default only in case of a $100\%$ default of the whole portfolio at $t = 0$. Finally, in order to obtain the $p$-th Position Thickness total amount, we multiply $THP_p(t)$ per the total outstanding balance $OBP(t)$ as
\begin{equation} \label{eq:th}
	TH_p(t) = OBP(t) \cdot THP_p(t)  
\end{equation}

\subsubsection{The PEAL Method Thickness definition} \label{Subsubsec: Step 9 - Thickness PEAL Method}
The Thickness computational method defined in the EU Regulation has a fatal flow: it only works in Design with horizontal Positions. As we have shown in the previous Sec. \ref{Sec: Step 6 - Designing}, the Positions' Designing can have also vertical Positions: therefore, it becomes mandatory to develop a more general approach in computing the Position Thickness that could be used in any Designing, irrespectively by its complexity. Since, to compute other Features like the Regulatory Capital of Equation \eqref{eq: rcny} we only need the $p$-th Position Thickness $TH_p(t)$, and since $THP_p(t)=OB_p(t)/OBP(t)$, then
\begin{equation} \label{eq: th}
	TH_p(t) = OB_p(t)
\end{equation}
Thus, the Thickness of the $x$-th Cost Position is
\begin{equation} \label{eq: thc}
	THC_x(t) = OB_x(t) = \sum_{t=0}^{TP} GC_x(t) - \sum_{\tau=0}^t GC_x(\tau)
	= \sum_{\tau=t+1}^{TP} GC_x(\tau)
\end{equation}
and the Thickness of the $y$-th Note Position is
\begin{equation} \label{eq: thn}
	THN_y(t) = OB_y(t) = \sum_{t=0}^{TP} GN_y(t) - \sum_{\tau=0}^t GN_y(\tau)
	= \sum_{\tau=t+1}^{TP} GN_y(\tau)
\end{equation}
and considering that we already have both the Gross Dimensioned Cost Position $GC_x(t)$ and the Gross Dimensioned Note Position $GN_y(t)$, we can compute their respective Thickness $THC_x(t)$ and $THN_y(t)$ without the need of their respective attachment points $AP_p(t)$ and detachment points $DP_p(t)$, simplifying the whole computation and model complexity, yet respecting the spirit of the European Regulations.

\subsection{Regulatory Capital (RC)} \label{Subsec: Step 9 - Regulatory Capital}
The Regulatory Capital (\textbf{RC}), or capital requirement, is the amount of capital a bank or other financial institutions must hold in relation to the riskiness of their assets as required by their financial regulators. Therefore, as explained in the previous Sec. \ref{Subsec: Step 5 - Positions Qualities}, the RC is mathematically irrelevant to define a Structuring Method, and it is only an external requirement of some type of investors. In fact, if a Position were held by an investor that is not required to compute the RC, this mere fact would not impact the securitization intrinsic characteristics. For this reason we explicitly said that the \textit{seniority} is just a \say{quality} $q$ that each Position obtains depending on its Designing and Gross Dimensioning, whose effect is only relevant when computing the Regulatory Capital. Indeed, given the same number of Positions $P$ and the same Gross Dimensioning, a diverse Designing would assign to each Position a different \textit{quality} $q$. 

Notice that the slice of $p$-th Position absorbing the Complementary Loss Tranche $CLT(t)$ gets the quality of Senior ($q = Senior$ or $q = SN$). The slice absorbing the Second Loss Tranche $SLT(t)$ gets the quality of Mezzanine ($q = Mezzanine$ or $q = MZ$) and the slice absorbing the First Loss Tranche $FLT(t)$ gets the quality of Junior ($q = Junior$ or $q = JR$). The Regulatory Capital of the $x$-th Cost and $y$-th Note Position is the $3$-rd Feature that shall be provided to the securitization stakeholders on an ongoing basis. 

As we explained in the previous Sec. \ref{Sec: Step 5}, nowadays the Risk Weight of the Cost Position is zero $RW^{(C)} = 0$ thus $RWC = 0$. Therefore, for the purpose of this Section, we will compute the Regulatory Capital only on the $Y$ Note Positions $RCN_y(t)$. The general equation to compute the Regulatory Capital of the $y$-th Note Position is 
\begin{equation} \label{eq: rcny}
	RCN_y(t)= \sum_{q} RCN_{(y,q)}(t) = RCN_{(y,SN)}(t) + RCN_{(y,MZ)}(t) + RCN_{(y,JR)}(t)
\end{equation}
where $y = 1 \dots Y$. The Note Position Regulatory Capital $RCN_{(y,q)}(t)$ for the quality $q$ is obtained by multiplying the risk-weighted asset $RWA_i$ of the $i$-th Gross Dimensioned Vertical Component $GV_i(t)$ that belongs to the $y$-th Note Position that absorbs the relative $q$ Tranche (see exemplified system of equations in Sec. \ref{Subsec: Step 7 - GC e GN}), by the its capital adequacy ratio (\textbf{CAR}, see CRR, Article 92) as
\begin{equation} \label{eq: rcnyq}
	RCN_{(y,q)}(t)= \sum_{i} RWA_{(i,q)}(t) \cdot CAR =  \sum_{i} (OB_i(t) \cdot RW_{(i,q)}) \cdot CAR
\end{equation}
where $OB_i(t)$ is Equation \eqref{eq: thn}. Notice that Reg 2401 goes into great detail in explaining how to compute the different $RW_{(i,q)}$ as a function of the relative Thickness and other parameters. We will not deep dive into such aspects, but we will postpone its detailing to a next paper. Therefore, if we consider the example in Sec. \ref{Subsec: Step 7 - GC e GN}, the $y$-th Note Position Regulatory Capital $RCN_y(t)$ can be computed with the following system of equations
\[
\begin{cases} \label{eq: rcnyt}
	RCN_1(t) = & OB_3(t) \cdot RW_{(3,SN)} \cdot CAR \\
	RCN_2(t) = & OB_5(t) \cdot RW_{(5,MZ)} \cdot CAR\\
	RCN_3(t) = & OB_7(t) \cdot RW_{(7,JR)} \cdot CAR\\
	RCN_4(t) = & [ OB_4(t) \cdot RW_{(4,SN)} + OB_6(t) \cdot RW_{(6,MZ)} + OB_8(t) \cdot RW_{(8,JR)} ] \cdot CAR\\
\end{cases}
\]
where $OB_i(t)$ and $RWA_{(i,q)}(t)$ are components of the function \eqref{eq: thn}. 

\subsection{Coefficient of Variation Adjusted (CVA)} \label{Subsec: Step 9 - CVA}
Let the $x$ Cost Coefficient of Variation Adjusted $CVAC_x(t)$ be ratio between the Gross Dimensioned Cost Position $GC_x(t)$ minus the 
expected deviation of the Net Dimensioned Cost Position $NC_x^{(s)}(t)$ and the $x$-th Cost Position's total Thickness $TH_x$ and as
\begin{equation} \label{eq: cvax}
	CVAC_x(t)=  \frac{ GC_x(t) - \langle NC_x^{(s)}(t)\rangle }{TH_x}
\end{equation}
Let the $y$ Note Coefficient of Variation Adjusted $CVAN_y(t)$ be ratio between the Gross Dimensioned Note Position $GN_y(t)$ minus the expected deviation of the Net Dimensioned Note Position $NN_y^{(s)}(t)$ and the $y$-th Note Position's total Thickness $TH_y$ and as
\begin{equation} \label{eq: cvay}
	CVAN_y(t)=  \frac{ GN_y(t) - \langle NN_y^{(s)}(t)\rangle }{TH_y}
\end{equation}
The Coefficient of Variation Adjusted is the $3$-rd Feature that shall be provided to the securitization stakeholders on an ongoing basis. Notice that the $CVA(t)$ of the $p$-th Positions absorbing the Complementary Loss Tranche $CLT(t)$ shall be systemically lower than those of the others, especially of the Positions absorbing the First Loss Tranche $FLT(t)$. If we have correctly applied all the PEAL Method requirements, then we know that a Structuring Method is not compliant with Reg 2402 if, by plotting all the Cost and Note CVA on the same graph, at any time $t$ the lines intersect. Indeed, if they do intersect it means that in $t$ a \say{higher-tier} Position is absorbing more risk than a \say{lower-tier} Position, resulting in a breach of Reg 2402 article 2(1)(b) provisions. The securitizations that do not respect this principle should not be allowed to be structured. In general, it is possible to avoid this kind of situation by having the Positions absorbing the $FLT$ paid bullet at $TP$. With this little shrewdness, the arranger is always sure that any Structuring Method respect at least the spirit of Reg 2402.      

\subsection{Fair Value (FV)} \label{Subsec: Step 9 - Fair Value}
The Fair Value $FV$ is the estimated price at which an asset, in our case the $y$-th Note Position, is bought or sold when both the buyer and seller freely agree on a price. The most accepted methodology to calculate an asset fair value is the so called \say{discounted cash flow} or \say{DCF}, where we actualize with a certain discount factor $d$ the future asset free cash flows $FCF$ to a certain date $\tau = t$. 
\begin{equation} \label{eq: dcf}
	DCF(t)= \sum_{\tau=t}^T \frac{FCF(\tau)}{(1+d_{\tau})^{\tau}}
\end{equation}
In the PEAL Method we compute the $DCF(t)$ on a set of representative Scenarios $S$ deriving a probability density function $DCF^{(s)}(t)$, and we compute the Fair Value as
\begin{equation} \label{eq: fvy}
	FVY_y(t)= \langle FVY_y^{(s)}(t) \rangle
\end{equation}
that is the expected value of the $S$ Scenarios' value where
\begin{equation} \label{eq: tv}
	FVY_y^{(s)}(t) = \sum_{\tau=t}^T \frac{NN_y^{(s)}(\tau)}{(1+\eta_{\tau})^{\tau}}
\end{equation}
where $NN_y^{(s)}(\tau) = FCF(\tau)$ because the Ned Dimensioned Note Position is already \say{net} of each Scenario $s$ losses; the discount factor $d = \eta$ as the annual expected inflation rate (e.g. $\eta = 2\% $) because if we were to discount the $FCF(t)$ for an higher rating the computation of the expected loss would lead to a double jeopardy; and $d_{\tau} = \eta_{\tau}$ as the monthly inflation rate (e.g. $\eta_{\tau} = \eta / 12$). The Fair Value is the $4$-th Feature that shall be provided to the securitization stakeholders on an ongoing basis. For each $y$-th Note Position in any Scenario $s$, the Fair Value is the $4$-th Feature that shall be provided to the securitization stakeholders on an ongoing basis. 

The Fair Value $FVY_y(t)$ corresponds to the price that investors of a neutral and liquid market, with perfect information, would pay to buy the $Y$ Notes. It is a very useful synthetic measure that the servicers must always provide to the sellers, especially because it gives them an instrument to evaluate the \textit{fairness} of a price. In fact, imputing the price offered by the buyer into the Fair Value's empirical cumulative distribution function, it is possible to derive the specific quantile to whom that price corresponds to: if the quantile is too far away from the average, it is an indication to the seller that, unless has a valid motive to do otherwise, shall avoid to sell rashly and wait for prices that are more aligned with the intrinsic riskiness of the Note Position. We believe that, in a rational market, the more information like this one is provided to the investors, the more the market is going to become liquid and efficient; to such an aim, the securitization servicers should provide this kind of information to the investors with high frequency and granularity. 

\subsection{Internal Rate of Return (IRR)} \label{Subsec: Step 9 - IRR}
Investors use a synthetic metric call the \say{internal rate of return} $irr$ to compute the annual rate of growth that an investment is expected to generate. Often this metric is used to compare different investments with similar standard deviation, and therefore is especially relevant as a \say{rule-of-thumb} measure. The following is the general equation
\begin{equation} \label{eq: irr}
	irr:\sum_{t=0}^{TP}\frac{ECF(t)}{\left(1+irr\right)^{t}}=C_0
\end{equation}
where $ECF(t)$ represents the expected asset cash flows, $C_0$ represents the price paid at time $t=0$ to buy the asset, and $TP$ represent the fact that the investors are willing to buy the $(K,N)$ Exposures cash flows up to the Note Position maximum duration. The Internal Rate of Return $irr$ is the $5$-th Feature that shall be provided to the securitization stakeholders on an ongoing basis. Let $girr_y$ be the gross $irr$ of the $y$-th Note Position as 
\begin{equation} \label{eq: girr}
	girr_y:\sum_{t=0}^{TP}\frac{GN_y(t)}{\left(1+girr_y\right)^{t}}=CY_{(0,y)}
\end{equation}
where $ECF(t)$ is substituted by the $y$-th Gross Dimensioned Note Position $GN_y(t)$, and where $CY_{(0,y)}$ represents the initial capital paid by the investors at time $t=0$ to buy the $y$-th Note Position. Let $nirr_y$ be the net $irr$ of the $y$-th Note Position as 
\begin{equation} \label{eq: nirr}
	nirr_y:\sum_{t=0}^{TP}\frac{NN_y(t)}{\left(1+nirr_y\right)^{t}}=CY_{(0,y)}
\end{equation}
where the asset expected cash flows $ECF(t)$ are substituted by the $y$-th Net Dimensioned Note Position cash flows $NN_y(t)$, and where $CY_{(0,y)}$ represents the initial capital paid by the investors at $t = 0$ to buy the $y$-th Note Position. Notice that $CY_{(0,y)}$ is
\begin{equation} \label{eq: CYy}
	CY_{(0,y)} = cpy_y \cdot CY_0
\end{equation}
where $cpy_y$ is a percentage of the total price that the investors would pay to buy all the $Y$ Note Positions at once $CY_0$, with $\sum_y cpy_y = 100\%$. Then 
\begin{equation} \label{eq: CY}
	CY_0 = min(C_0,FVY_y(0))
\end{equation}
is the minimum between the initial $N$ Exposures total capital $C_0$ and the total Note Position Fair Value where the mean of the previous Equation \eqref{eq: fvy} is substitute by the $\sum_y FVY^{(s)}_y(t)$. In general terms, $CY_0 = FVY(0)$ oftentimes when the Exposures are composed by Negative Event Exposures, in which the expected cash flows $ECF(t)$ are on average lower then the initial capital $C_0$, and thus the portfolio is sold at discount.  

By considering the previous Equation \eqref{eq: CYy}, we can infer that the number $Y$ of Note Positions do have a great relevance in computing the various internal rate of return. Indeed, if $Y = 1$ then $CY_{(0,1)} = CY_0$ and thus there is only \say{one} potential $girr$ and $nirr$. When $Y = 2$, then $CY_{(0,1)} = cpy_1 \cdot CY_0$ and $CY_{(0,2)} = (1 - cpy_1) \cdot CY_0$, generating an infinity of $girr$ and $nirr$ solutions in correspondence with the segment $CY_{(0,1)}+CY_{(0,2)}=CY_0$, $CY_{(0,1)},CY_{(0,2)}\geq0$. When $Y > 2$, then the infinities of $girr$ and $nirr$ solutions correspond to the intersection of the $\sum_y CY_{(0,y)}=CY_0$ plane with the positive orthant $CY_{(0,y)}\geq0$. It follows that, whilst the PEAL Method can determine the optimal Gross and Net cash flows to the $x$-th Cost and $y$-th Note Position, it does not determine the $girr$ and the $nirr$. Indeed, the $CY_{(t,y)}$ definition is not an intrinsic characteristic of any securitization, but rather a decision that is totally market-driven, a choice done to meet the risk appetites of different investors. 
As $Y \geq 2$, the number of potential $girr$ and $nirr$ solutions becomes infinite. Therefore, it becomes imperative to explore reasonable constraints to narrow down these solutions to a manageable subset. This is especially crucial in alignment with the spirit of securitization regulations, which are primarily aimed at safeguarding investors and enhancing transparency. Investigating such constraints is essential to uphold investor protection and promote market integrity.

\section{Step 10: Optimize the Structuring Method} \label{Sec: Step 10}
The 10th Step of any Structuring Method is to optimize the Features of Step 9, by changing the $X$ Cost and $Y$ Note Positions Designing of Step 6 and Gross Dimensioning of Step 7 or by adding the optimization inbound building-block $z(t)$. Currently, Equation \eqref{eq: Virtual Position} has been set to be sure that the $X$ Cost and $Y$ Note Positions Designing of Step 6 and Gross Dimensioning of Step 7 are optimized for the Regulatory Capital calculation. In the end, there is no right or wrong optimization, since all depends on the arranger or the investors' needs. The only \textit{vinculum} is to always pay attention to the Features' results respect, on an ongoing basis, the Reg 2402 provisions.

\section{Next steps}\label{Sec: Next Stpes}
The exploration of various opportunities for future investigation offers valuable insights into the complex landscape of securitization practices. 

One such avenue involves standardising the modeling of the 11 different Types of Exposures outlined in steps 1 to 4, utilizing foundational equations provided in Sec. \ref{Sec: C1 - unboxing}. This systematic analysis would allow both to shed light on the Features and risk management considerations associated with each Type, thereby informing decision-making processes and enhancing risk assessment practices within securitization transactions, and to produce standardised \say{templates} for new securitizations.

Another area that needs to be assessed is the evaluation of the impact of Euribor changes on Note Positions risk/return profiles: an accurate modelling of the Euribor dynamics is needed to understand whether resulting excess losses exceed excess interest paid to investors. Furthermore, investigating the potential significance of using excess returns to mitigate losses on \say{higher-tier} Positions could provide valuable insights into effective risk management strategies within securitization structures.

Future research endeavors could also delve into alternative approaches to compute the optimization inbound building-block $z(t)$. By exploring different methodologies, we can deepen our understanding of optimization techniques and potentially uncover more efficient processes.

Another avenue for further investigation might be the exploration of various Position Designs and their respective Waterfall Configurations. Diving into different designs and configurations promises to enrich our understanding of the securitization process and its implications for risk management and feature computation.

Additionally, future research should explore the possibility of designing Waterfall Configurations that eliminate the so-called \say{Junior} Note, thereby structuring risk-intensive securitizations that may otherwise be deemed unfeasible due to excessive regulatory capital costs.

Further exploration is needed to define optimal frequencies in different Waterfall Configurations. This involves assessing the impact of relaxing one or more of the \say{3-frequency} rules presented previously and evaluating how these relaxations may influence risk management strategies. By investigating alternative configurations, we can contribute to the development of more robust securitization practices.

Another area ripe for exploration revolves around incorporating advanced risk weight calculation methods into the regulatory capital computation. By understanding the relationship between risk weights and Position Thickness, we can improve the precision and reliability of risk weight calculations, thereby enhancing risk management practices.

Furthermore, the development of constraint optimization models tailored for $irr$ computation in securitization processes is essential. These models should optimize $irr$ while satisfying regulatory requirements, investor preferences, and other relevant constraints, to enhance decision-making processes.

Clustering different Structuring Methods based on the outcome of their Features offers another avenue for exploration. This approach provides stakeholders with valuable insights into anticipated ongoing securitization performance, fostering the creation of shared best practices to expedite the structuring process without compromising soundness.

Moreover, developing ongoing management Key Performance Indicators to compare historic expected, historic real, and forward-looking real expected Scenarios offers valuable insights into actual securitization performance. Extending the methodology to ongoing management enhances our ability to monitor and optimize securitization performance throughout its life cycle.

Finally, estimating the time of risk hedging for each Position is crucial. Determining the date from which the cash Advances absorb the position losses, thus making the Position virtually risk free, allows us to manage risk effectively and adjust regulatory capital requirements accordingly.

In conclusion, the areas identified in this paper present promising avenues for further exploration, deepening our understanding of securitization mechanisms and contribute to the development of more effective Structuring Methods, risk management strategies, enhancing regulatory supervision.

\section{Conclusions}\label{Sec: C7 - Conclusion}
This paper introduces the PEAL Method, a comprehensive 10-step mathematical framework designed to facilitate the structuring of securitizations across various Designing, including those with vertical Positions. It offers a universal approach applicable to any combination of Type of Exposures and Waterfall Configurations, positioning existing securitization structures as specific cases within our general framework. Moreover, it encompasses a comprehensive analysis across all time periods.

Our framework meticulously models the distribution of Exposures inbound cash flows from assets to outbound cash flows across positions. By providing a structured approach, we clarify payment priorities and enhance transparency regarding ongoing risk characterization for both the asset and liability sides of securitizations.

Moreover, our structured methodology encourages the explicit delineation of all relevant hypotheses, formulas, models, and algorithms used in the structuring phase. By incorporating these details into legal documentation, practitioners can furnish investors and supervisory authorities with comprehensive information necessary for replication and evaluation. This enhancement mirrors the transparency and reproducibility standards of scientific papers, enabling stakeholders to scrutinize underlying assumptions and test hypotheses reflected in legal contracts. Ultimately, this approach promises to simplify comparisons among different securitizations, foster transparency, and facilitate regulatory supervision, thereby mitigating potential disputes among stakeholders.

To further bolster stakeholder risk assessment, we advocate for the provision of a minimum set of features to investors on an ongoing basis. These metrics empower investors to gauge the risk characterization of positions more accurately, leading to pricing convergence closer to fair values. Consequently, our framework promotes standardization and cross-comparability among diverse structuring methods, augmenting market transparency and potentially curbing structuring and due diligence costs over time.

However, caution is warranted when employing complex designs, as they may obscure underlying mathematical intricacies and potentially disadvantage investors if not properly communicated.

In summary, this foundation paper proposes a novel securitization framework, offering a comprehensive methodology to design, evaluate, and compare different securitizations over time. Our approach facilitates the clustering of structuring methods based on expected features, providing stakeholders with valuable insights into anticipated ongoing securitization performance. By fostering the creation of best practices and promoting regulatory compliance without compromising soundness, our framework aims to contribute to the evolution and improvement of securitization practices.


\newpage

\appendix
\section{Description of the elements of any List of Events} \label{Appendix: List of Events}
This is a non-exhaustive list of the main Events by Type of Exposures:
\begin{table}[ht] 
	\begin{center}
		\begin{tabular}{ |c|c|c|c|c|c|c|}
			\hline
			\textbf{TE} & \textbf{LE} & \textbf{NE} & \textbf{EX}\\
			\hline
			\hline
			QP & pe, dh & 2 & tm, oi  \\ 
			QS & pe, dh, npd, jl & 4 & tm, oi  \\
			CL & pe, de, eu, trl & 4 & tm  \\ 
			EA & imp, fr, ts, sp & 4 & en  \\
			NE & fr, trl, rr, tr & 4 & tm \\
			EE & npd, fr, ts, sp & 4 & tm, en  \\
			CC & de, fr, eu, trl & 4 & tm, oi \\ 
			SL & pe, de, dh, eu, trl & 5 & tm \\
			ML & pe, de, cd, eu, trl & 5 & tm  \\ 
			AL & pe, de, cd, eu, trl & 5 & tm  \\
			RE & npd, fr, cd, ts, sp, nr & 6 & tm, en  \\
			\hline
		\end{tabular}
	\end{center}
	\caption{List of Events by Type. \textbf{TE} stands for Type of Exposures; \textbf{LE} stands for List of Events); \textbf{NE} is the Number of elements in the List of Events; \textbf{EX} stands for Extreme Events, that are so rare that, unless there is a compelling reason to use them, they shall never be included in the LE $\Lambda$. \label{Tab: List of Events}} 
\end{table}

\noindent Hereby, a brief description of each Event of the previous Table \ref{Tab: List of Events}:
\begin{itemize}
	\item cd: stands for Collateral Depreciation, describes the negative Event where the securitization collateral diminishes in value versus its time of structuring $\hat{\tau}_k$;
	\item de: stands for Default Event, describes the negative Event where a borrower does not repay their obligations, and it is assumed that does not come back to life;
	\item dh: stands for DeatH, describes the negative Event where the borrower dies;
	\item en: stands for Extreme Natures, describes Events that can impair the collaterals;
	\item eu: stands for EUribor, describes a hybrid Event;
	\item fr: stands for FRaud, describes the negative Event where there exists some kind of fraud (the collateral does not exist, the credit card has been stolen, etc);
	\item imp: stands for IMPairment, describes the negative Event where a perishable asset is no more available to be sold or does not exist anymore; 
	\item jl: stands for Job Loss, describes the negative Event where an individual backing the Exposure loses their job, impairing the Exposure's capacity to repay its obligations;
	\item npd: stands for Not PaiD, describes some possible negative Events (e.g. not being paid by an employer, renter not paying, etc);
	\item nr: stands for Not Rented, describes a hybrid Event; 
	\item oi: stands for Over-Indebtedness, describes the negative Event where a judge rules the debt inside the securitization be noncollectable in part or in whole; 
	\item pe: stands for Prepayment Event, describes the negative Event where a borrower decides voluntarily to pay back the capital before its end $T_{k,n}$;
	\item rr: stands for Recovery Rate, describes a hybrid Event;
	\item sp: stands for Selling Price, describes a hybrid Event;
	\item tm: stands for Moratorium Time, describes the negative Event where a State may lawfully impose to accept the postponement of installments by $t_m$ months;  
	\item trl: stands for Return to Life Time, describes the positive Event where a Negative Event Exposure starts to repay its debt; 
	\item tr: stands for Recovery Time, describes a hybrid Event;
	\item ts: stands for Selling Time, describes a hybrid Event.
\end{itemize}
Notice that a hybrid Event changes, in positive or negative, vs its Base Input value at $\hat{\tau}_k$ of structuring. The mathematical description of each one of the asset $A(t)$ and event recovery $E(t)$ equations should be defined in future research papers. 

\end{document}